\documentclass[%
 reprint,
 superscriptaddress,
 amsmath,amssymb,
 aps,
prb,
]{revtex4-2}
\usepackage{xcolor}
\usepackage{graphicx}
\usepackage{dcolumn}
\usepackage{bm}
\usepackage{soul}

\begin{document}
\preprint{APS/123-QED}

\title{Control of a ferrimagnet phase by a two-component magnetic field}

\author{D.O. Ignatyeva}
\email[]{ignatyeva@physics.msu.ru}
\affiliation{Faculty of Physics, Lomonosov Moscow State University, Leninskie gori, 119991 Moscow, Russia}
\affiliation{Russian Quantum Center, 121205 Moscow, Russia}

\author{N.A. Gusev}
\affiliation{Russian Quantum Center, 121205 Moscow, Russia}

\author{A.K. Zvezdin}
\affiliation{Russian Quantum Center, 121205 Moscow, Russia}
\affiliation{Prokhorov General Physics Institute of the Russian Academy of Sciences, 119991, Moscow, Russia}

\author{V.I. Belotelov}
\affiliation{Faculty of Physics, Lomonosov Moscow State University, Leninskie gori, 119991 Moscow, Russia}
\affiliation{Russian Quantum Center, 121205 Moscow, Russia}

\date{\today}

\begin{abstract}
We report a theoretical study of the phase diagram of a ferrimagnetic iron-garnet with uniaxial anisotropy near a magnetization compensation point in the presence of a two-component magnetic field. The study is performed based on a quasi-antiferromagnetic approximation. The number and stability of the equilibrium states of the Neel vector are analyzed using the effective energy function. It is shown that application of the small out-of-plane magnetic field in addition to the stronger in-plane magnetic field significantly changes the equilibrium states of a ferrimagnet. The possibilities to control the equilibrium Neel vector position and to switch between the monostable and bistable states by tuning the value and ratio of the in-plane and out-of-plane magnetic field components are demonstrated. This opens new possibilities for the utilization of ferrimagnets since the magnetic field could be changed much faster than the temperature.
\end{abstract}

\maketitle

\section{Introduction}

Ferrimagnets are of prime interest among the different magnetically ordered materials as a material platform for various optomagnetic~\cite{kimel2019writing,stupakiewicz2017ultrafast, ignatyeva2024optical} and spintronic devices~\cite{zhang2023ferrimagnets, kim2022ferrimagnetic,bello2022field, chernov2020all}. Ferrimagnets are formed by several magnetic sublattices, which allows for tuning of their magnetic and magneto-optical properties by composition~\cite{bayaraa2019tuning}. The sublattice magnetic moments vary with temperature at different rates, therefore, the temperature dependence of the total net magnetization of ferrimagnets can be engineered on demand. In particular, magnetization compensation can be achieved~\cite{smart1955neel}. Inequivalence of the ferrimagnetic sublattices provides a possibility to control their magnetic properties via an external magnetic field~\cite{goranskii1969turning} and to detect the ferrimagnet state by magneto-optical methods, which is a high contrast to antiferromagnets.

Ferrimagnets with a compensation temperature attract much interest due to their peculiar features of static and dynamic magnetic properties. For example, strong variations of the domain structure~\cite{kalashnikova2012magneto, kuila2019moke, sharipov2019domain} and domain wall motion~\cite{logunov2021domain, kim2017fast}, as well as skyrmion formation~\cite{caretta2018fast} were recently demonstrated in the vicinity of this point. One of the most intriguing feature of a ferrimagnet is an existence of the collinear and non-collinear magnetic phases near the compensation point~\cite{zvezdin1995field,davydova2019h,suslov2023non} and an ability to realize the phase transition between the different ferrimagnetic phases~\cite{sabdenov2017magnetic}. 

Recent interest to ferrimagnets is due to their unusual ultrafast response to femtosecond laser pulses: unconventional dynamics across the compensation point~\cite{stanciu2006ultrafast,binder2006magnetization}, in the non-collinear phase~\cite{krichevsky2023unconventional} or near spin-flop transition~\cite{becker2017ultrafast}, magnetization reversal~\cite{stanciu2007subpicosecond, radu2011transient, vahaplar2009ultrafast, vahaplar2012all}, 
ultrafast laser-induced heating with a consequent magnetization reversal~\cite{deb2018controlling} or precession~\cite{dolgikh2022spin} and peculiar magnetization precession frequency dependence~\cite{stupakiewicz2021ultrafast,blank2021thz, mashkovich2022terahertz}.

Ultrafast response of a ferrimagnet strongly depends on its phase, i.e. the sublattice magnetization orientations and canting~\cite{krichevsky2023unconventional,pogrebna2019high,dolgikh2023ultrafast}. For example, it was shown that the dynamics in the non-collinear phase, including the frequencies of the two spin modes, significantly differs from the collinear one~\cite{krichevsky2023unconventional}. Thus, it is important to analyze the features of the ferrimagnet phase diagram and to obtain a possibility to control it. The straightforward way of ferrimagnet state control is to select a desired $(H,T)$ point since the ferrimagnetic phase in sensitive to both the temperature and the applied external magnetic field~\cite{clark1968neel}. Usually, quite large changes of both parameters are required. In the present manuscript we analyze how the ferrimagnetic phase diagram changes under application of the two-component magnetic field. We demonstrate that the small out-of-plane field can be used to efficiently control the state of a ferrimagnet, which opens new possibilities for spin dynamics control.

\begin{figure*}[ht]
\centering
(a)~~~~~~~~~~~~~~~~~~~~~~~~~~~~~~~~~~~~~~~~~~~~~(b)~~~~~~~~~~~~~~~~~~~~~~~~~~~~~~~~~~~~~~~~~~~~~~~(c)\\
\noindent{
\includegraphics[width=0.6\columnwidth]{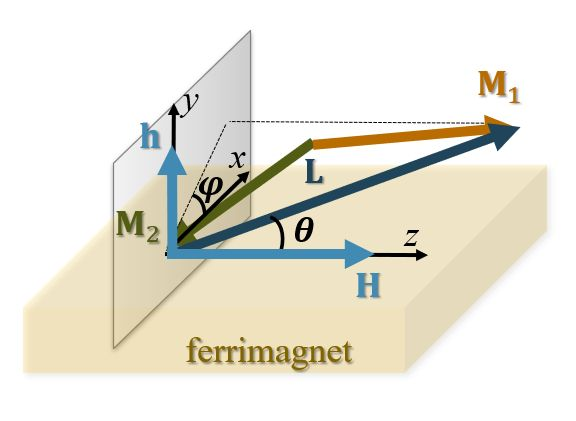}~~~~~
\includegraphics[width=0.5\columnwidth]{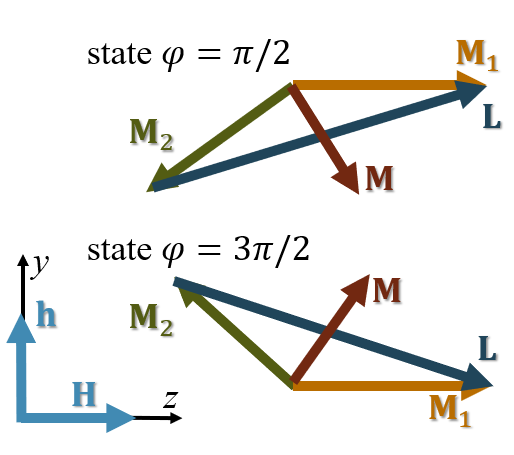}~~~~~
\includegraphics[width=0.8\columnwidth]{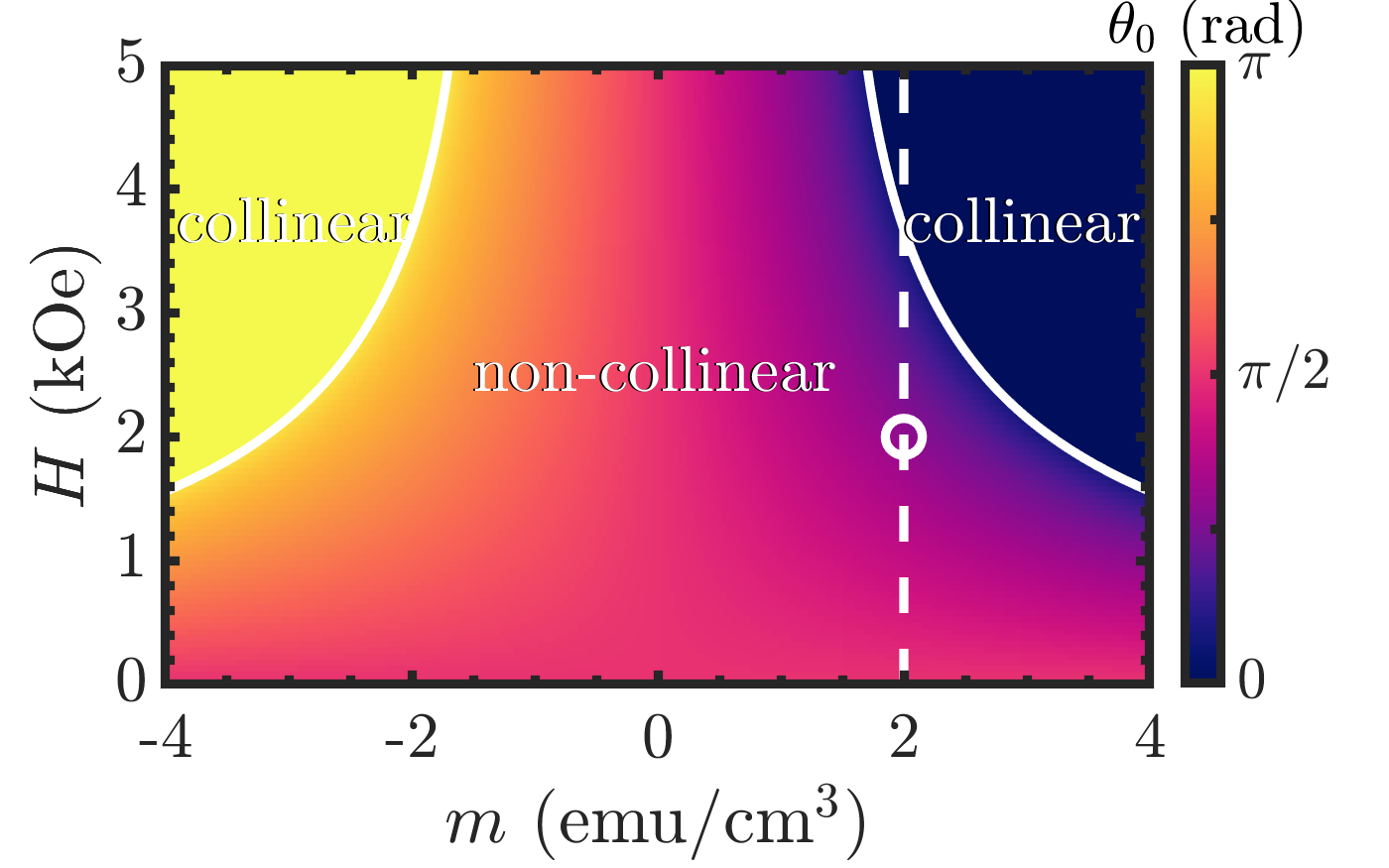}}\\
\caption{(a) Scheme of the considered configuration of a ferrimagnet. The angle between the $\mathbf{M_{1,2}}$ and $\mathbf{L}$ vectors is depicted exaggeratedly large for the sake of readibility. (b) Scheme of the two magnetic states with $\phi=\pi/2$ and $\phi=3\pi/2$ and similar $\theta_0$ angle of $\mathbf{L}$. Such states are degenerate for $h=0$. (c) Phase diagram of a uniaxial ferrimagnet with applied in-plane magnetic field $H$ ($h=0$). White solid lines correspond to the phase transition between the phases. White dashed line denotes the region that is studied in details further. White circle denotes the point ($m=2$~$\mathrm{emu/cm^3}$, $H=2$~kOe) that is analyzed in more details.}
\label{Fig: Scheme and PD for h=0}
\end{figure*}

The present work is devoted to the study of an impact of the two-component magnetic field on the magnetization states of a ferrimagnetic iron-garnet film with high uniaxial anisotropy in the vicinity of the compensation point. We show that small out-of-plane external magnetic field can be used for the efficient control of the equilibrium orientation of the sublattice magnetizations, the stability of the states and the whole phase diagram, as well. The paper is organized as follows. Section II is devoted to the basis of the theoretical analysis of the ferrimagnet magnetization states based on the quasi-antiferromagnetic approximation.  Section III is devoted on the qualitative analysis of the Neel vector orientations and trajectoris for different varitations of the parameters of the system. Analysis of the energy function of a ferrimagnet placed in the incline (two-component) magnetic field in quasi-antiferromagnetic approximation in terms of the Neel vector is provided in Section IV. We study the number and depth of the energy function minima that determine the equilibrium magnetization positions. In Section V we analyze how the equilibrium positions of the Neel vector change depending on the value of the out-of-plane magnetic field component. The most important impact of such a field is lifting of the degeneracy between the 'up' and 'down' states. Consequent variations of the bistability regions of the phase diagram are discussed in Section VI. With an increase of the out-of-plane component the bistability region shrinks. At the same time, as it is demonstrated in Section VII, the phase of the ferrimagnet is non-collinear in the whole range of magnetization and external magnetic field values considered. Thus, an ability to control the ferrimagnet phase and to switch between the monostable and bistable regimes by application of a small out-of-plane magnetic field is demonstrated.

\section{Theoretical description of a ferrimagnet in the incline magnetic field}

The ferrimagnets are characterized by magnetization compensation point, i.e. the temperature $T_\mathrm{M}$ at which the sublattice magnetizations compensate each other and, therefore, the net magnetization becomes zero. In the vicinity of $T_\mathrm{M}$, the sublattice magnetizations are nearly opposite to each other, except for a tiny canting angle and a small difference between their magnitude~\cite{tiablikov2013methods, gusev1959theory, zvezdin1972some}. Thus, for the description and analysis of the ferrimagnetic equilibrium states near the compensation point, it is convenient to use a quasi-antiferromagnetic approximation~\cite{blank2021thz, davydova2019ultrafast, krichevsky2023unconventional} rather than the sigma model~\cite{kim2017fast, ivanov2019ultrafast}. Quasi-antiferromagnetic approximation of ferrimagnet spin dynamics~\cite{blank2021thz, davydova2019ultrafast, krichevsky2023unconventional} is based on the assumption of nearly the same values of the sublattice magnetizations and small canting angles between them.

Let us consider a ferrimagnetic film of a two-sublattice ferrimagnet with $\mathbf{M_1}$ and $\mathbf{M_2}$ sublattice magnetization vectors. The film has uniaxial magnetic anisotropy with the axis normal to the film. The sublattice $\mathbf{M_{1,2}}$ vectors can be described by their magnitudes $M_{1,2}$ and angles $\theta_{1,2}, \phi_{1,2}$ determined in the spherical coordinate system with $z$-axis oriented in the film plane and $y$-axis along the anisotropy axis (see Fig.~\ref{Fig: Scheme and PD for h=0}a). As it was mentioned above, $\mathbf{M_1}$ and $\mathbf{M_2}$ have nearly the same values, but close to the opposite orientations near the compensation point, thus $M_1-M_2=m \ll M_1 + M_2$. This allows us to introduce an antiferromagnetic, or Neel, vector $\mathrm{L}=\mathbf{M_1}-\mathbf{M_2}$ described by the angles $\theta$ and $\phi$, so that:
\begin{eqnarray}
    \theta_1 = \theta - \varepsilon,~\theta_2 = \pi- \theta - \varepsilon , \\
    \phi_1 = \phi + \beta,~\phi_2 = \pi + \phi - \beta,
\end{eqnarray}
where angles $\pm\varepsilon$ and $\pm\beta$ give the canting of the sublattice magnetizations from $\mathbf{L}$ vector, $\varepsilon \ll \theta$, $\beta \ll \varphi$~\cite{zvezdin1979dynamics}.

\begin{figure*}[ht]
\centering
(a)~~~~~~~~~~~~~~~~~~~~~~~~~~~~~~~~~~~~~~~~~~~~~~~~~~(b)~~~~~~~~~~~~~~~~~~~~~~~~~~~~~~~~~~~~~~~~~~~~~~~~~~(c)\\
\includegraphics[width=0.6\columnwidth]{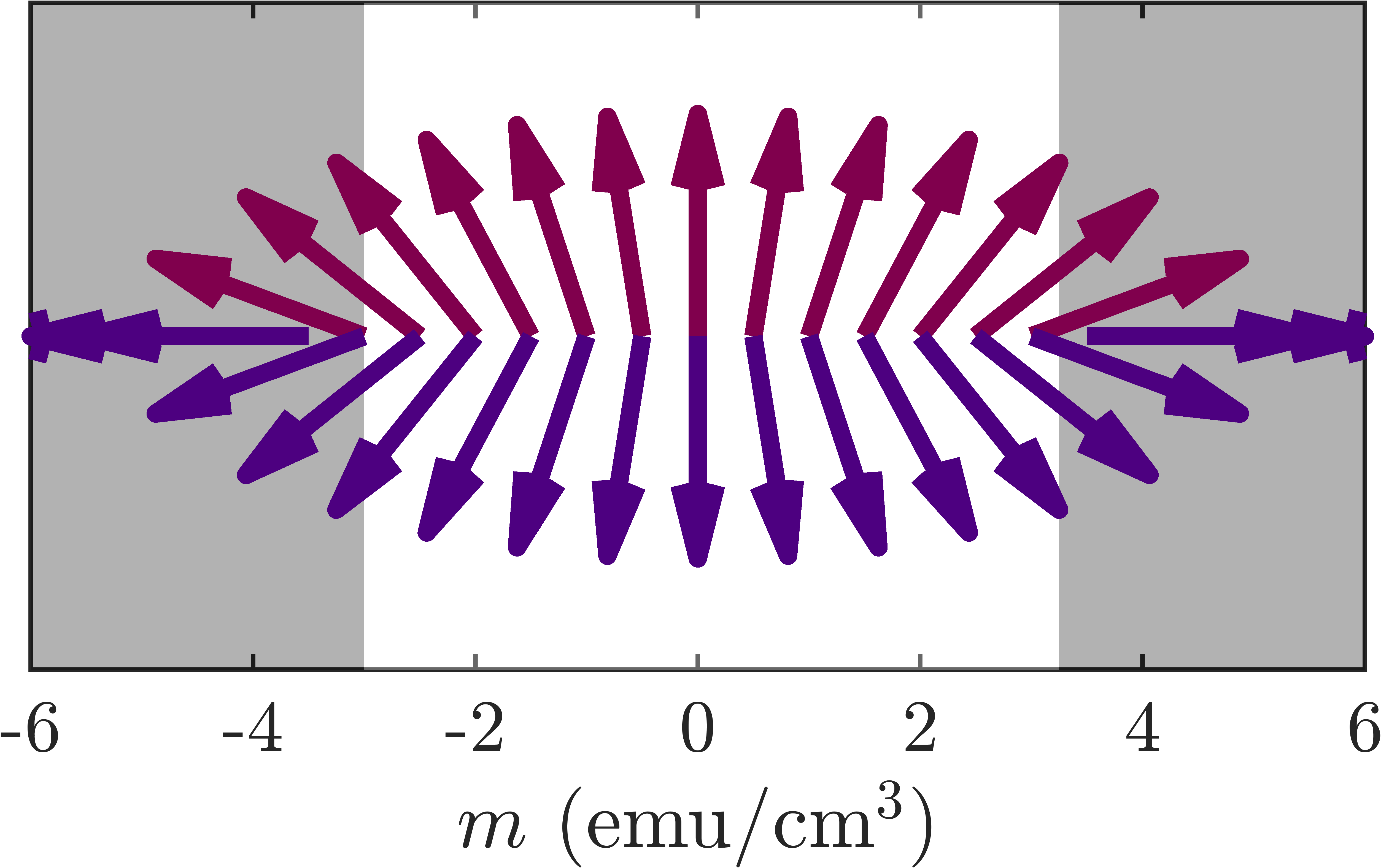}~~~~~~
\includegraphics[width=0.6\columnwidth]{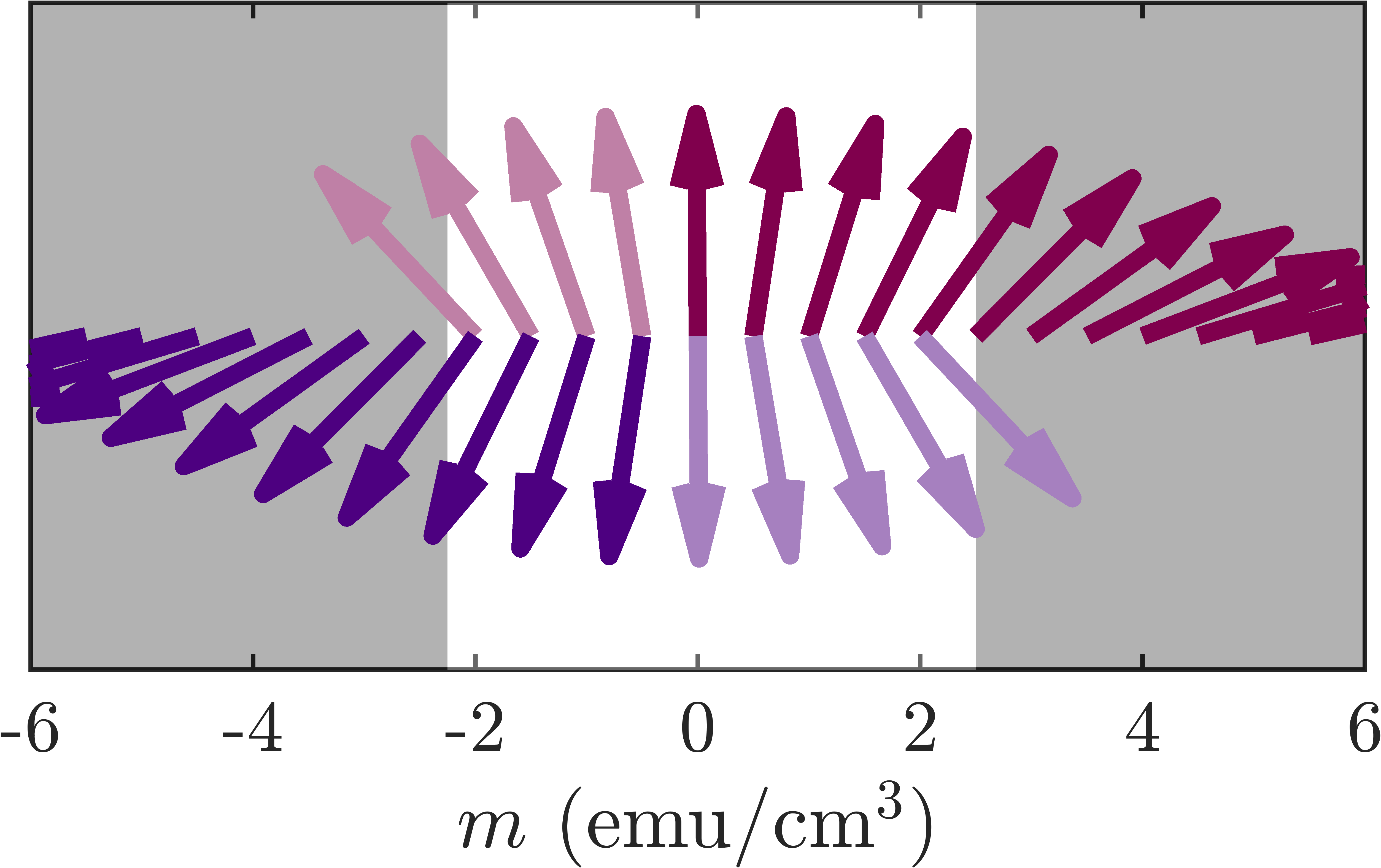}~~~~~~
\includegraphics[width=0.6\columnwidth]{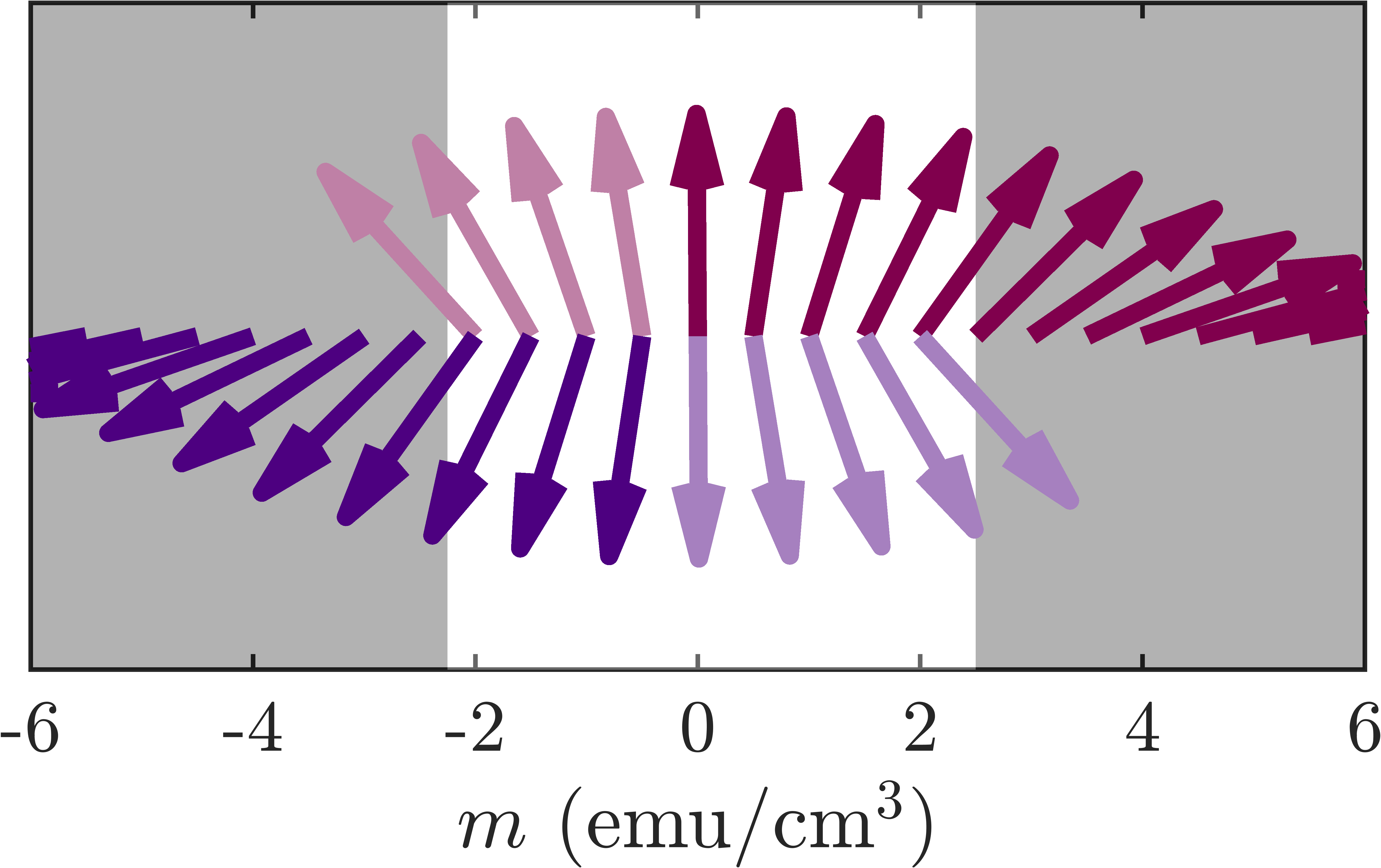}\\
(d)~~~~~~~~~~~~~~~~~~~~~~~~~~~~~~~~~~~~~~~~~~~~~~~~~~(e)~~~~~~~~~~~~~~~~~~~~~~~~~~~~~~~~~~~~~~~~~~~~~~~~~~(f)\\
\includegraphics[width=0.6\columnwidth]{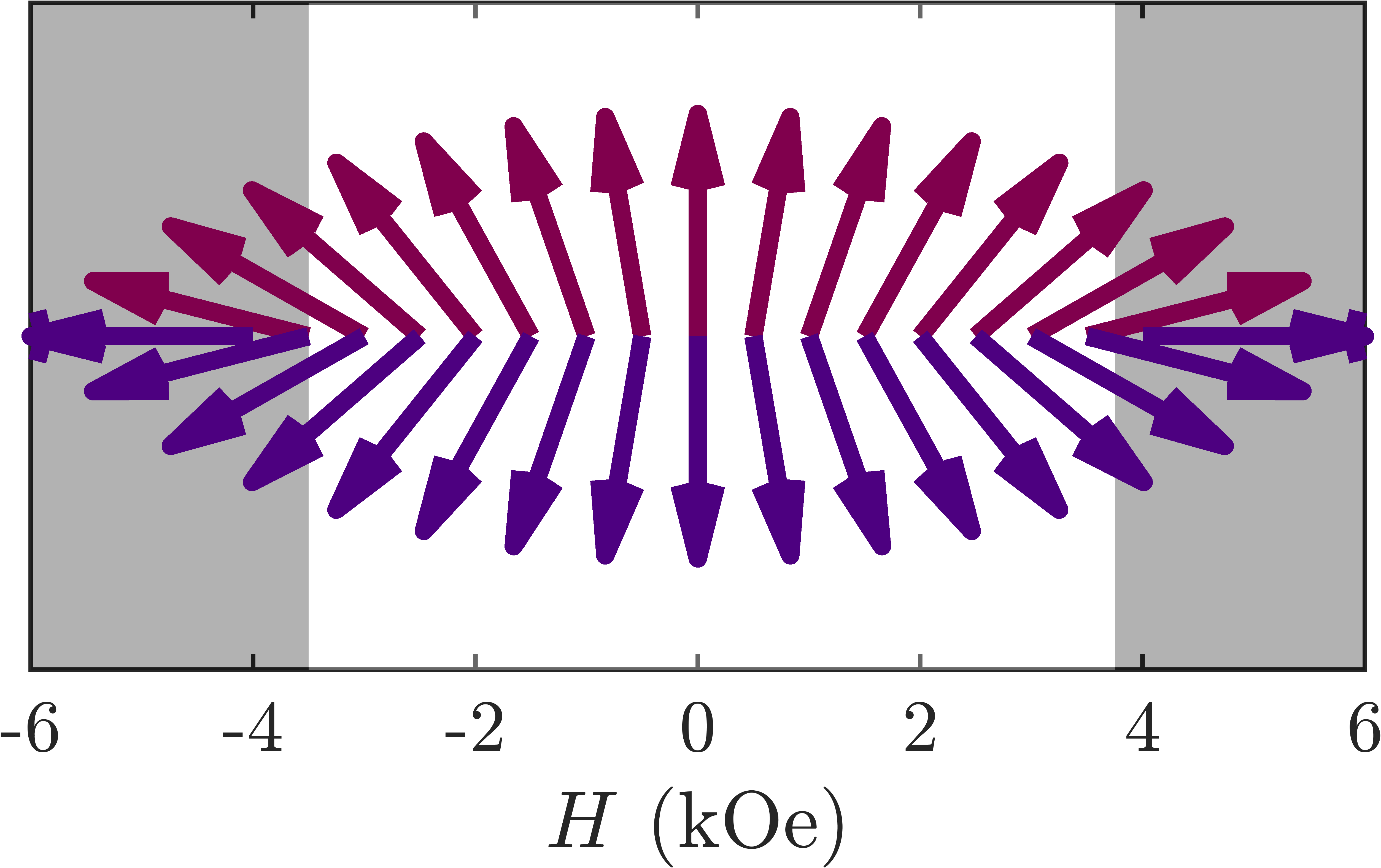}~~~~~~
\includegraphics[width=0.6\columnwidth]{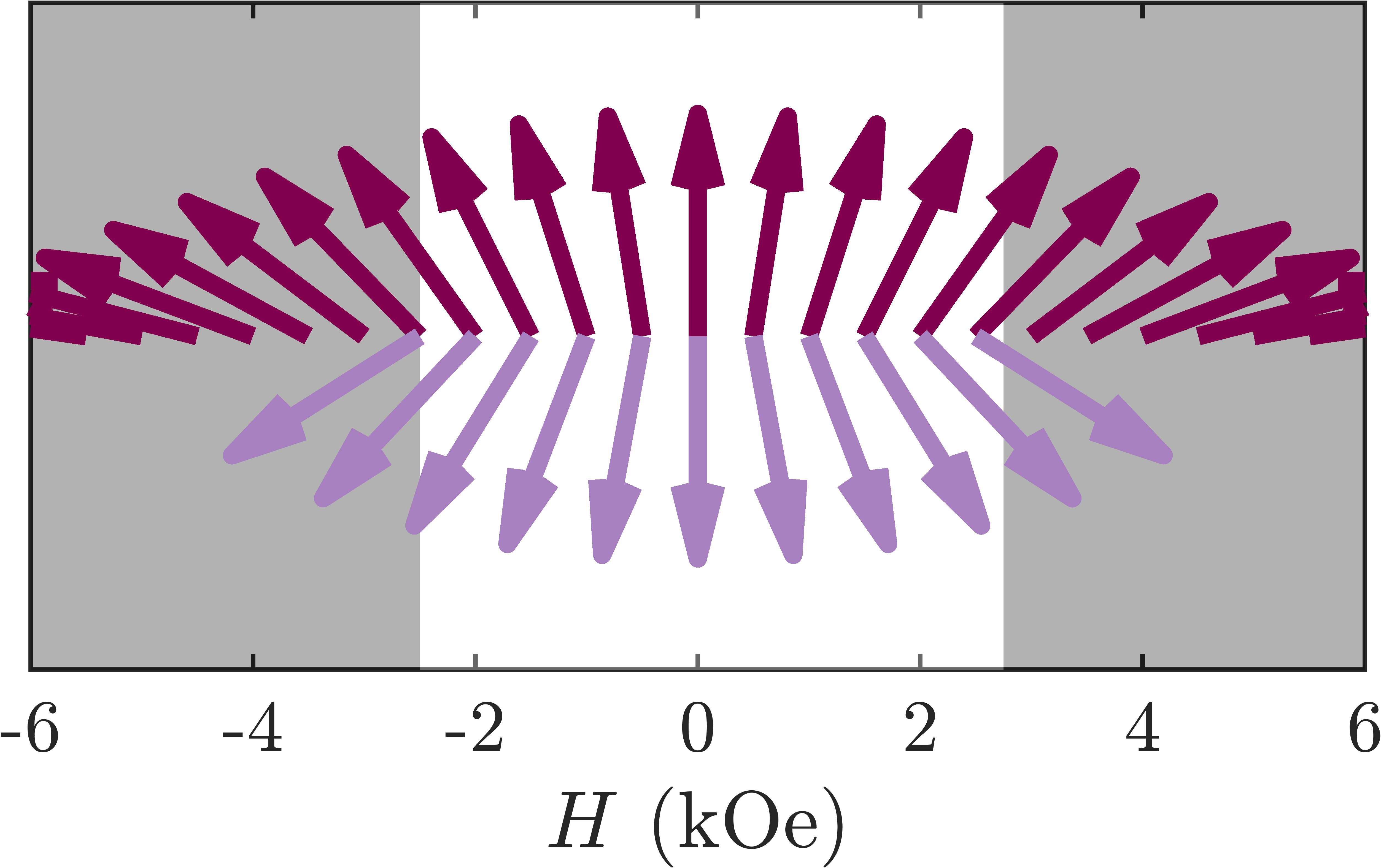}~~~~~~
\includegraphics[width=0.6\columnwidth]{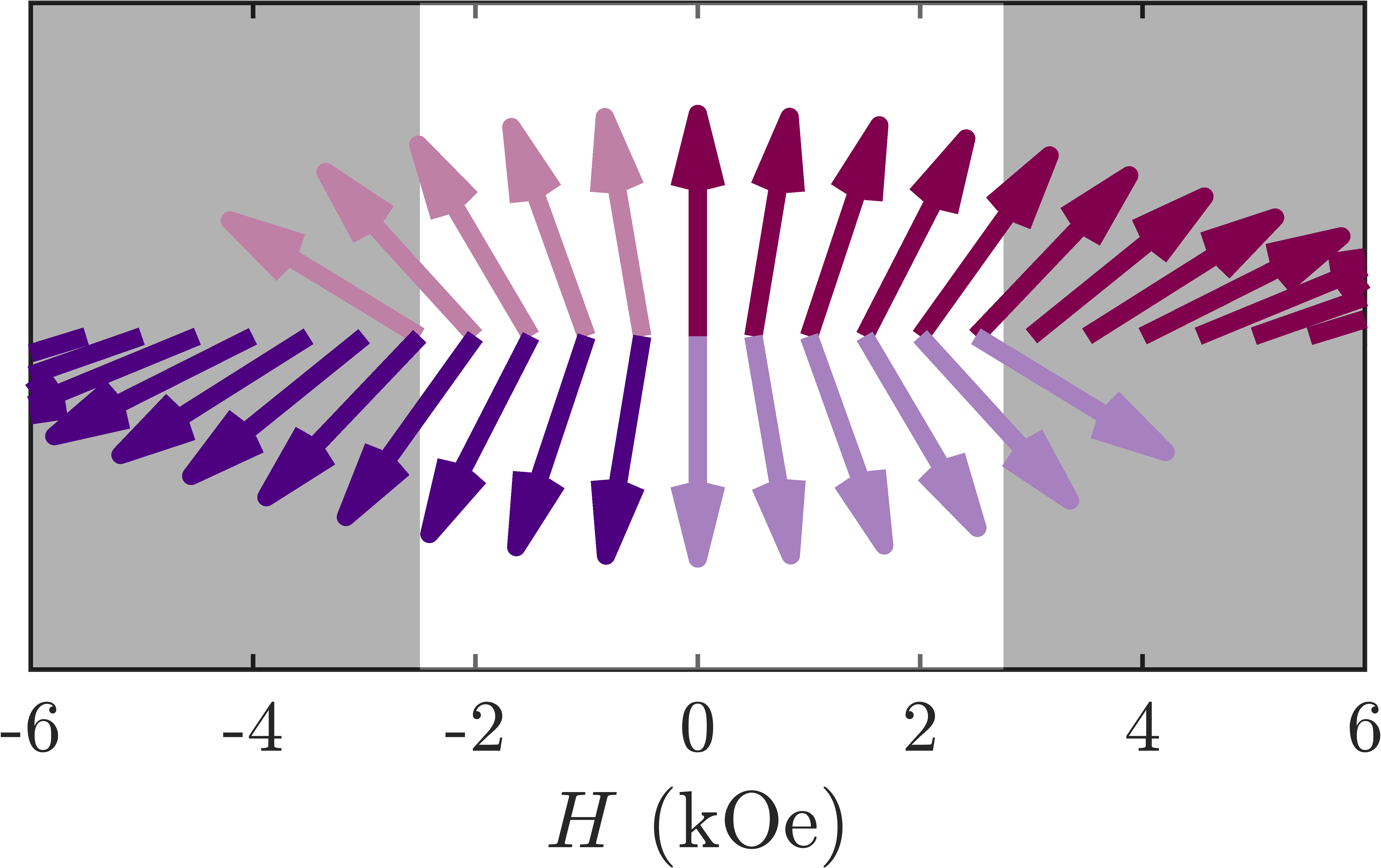}\\
\caption{Changes in the Neel vector $\mathbf{L}$ orientation for variation of the magnetization at the fixed $H=0.2$~kOe (top panel) or variation of the in-plane magnetic field $H$ at the fixed $m=2~\mathrm{emu/cm^3}$ (bottom panel) for (a,d) strictly in-plane ($h=0$) magnetic field, (b,e) in-plane magnetic field with an addition of a constant out-of-plane $h=0.2$~kOe magnetic field, (c,f) tilted external magnetic field with $h$ and $H$ simultaneously changing. Configuration (b,e) corresponds to the two-component magnetic field created by the two independent magnets, while (c,f) corresponds to the magnetic field of a single electromagnetic applied at an angle $\alpha=5^\circ$ to the film plane, so that $h=H \tan{5^\circ}$. The color of the arrows denotes the $\phi_0=\pi/2$ (dark and light pink color) and $\phi_0=3\pi/2$ (dark and light violet color) states. Dark pink and dark violet arrows show the states corresponding to the deepest minima of the effective energy function. Light pink and light violet arrows show the orientations corresponding to the less pronounced minimum of the effective energy function. Grey areas denote the regions with a single $\mathbf{L}$ equilibrium position, white region corresponds to the bistability region.}
\label{Fig: L arrows}
\end{figure*}

We consider a configuration where the magnetic field with a predominant in-plane component $\mathbf{H}$ along $z$ axis and a relatively small out-of-plane component $\mathbf{h}$ along $y$ axis is applied to the ferrimagnetic film. The projections of these $\mathbf{H}$ and $\mathbf{h}$ vectors on the corresponding $z$ or $y$ axis are denoted as $H$ and $h$, so that they both can have positive and negative values. 

The potential energy of the two-sublattice system in this case can be described as a sum of Zeeman, exchange and uniaxial anisotropy energy:
\begin{eqnarray}
    \Phi = -(\mathbf{M_1}+\mathbf{M_2})(\mathbf{H}+\mathbf{h})+2\Lambda\mathbf{M_1}\mathbf{M_2}-\nonumber\\
    -K_1\frac{(\mathbf{M_1}\mathbf{n})^2}{M_1^2}-K_2\frac{(\mathbf{M_2}\mathbf{n})^2}{M_2^2}, \label{Eq Phi gen}
\end{eqnarray}
where $\Lambda>0$ is a Weiss constant, $K_{1,2}$ are the anisotropy constants for each of the sublattices, $\mathbf{n}$ is anisotropy axis direction coinciding with the film normal. 

Using the potential energy in the form~\eqref{Eq Phi gen}, and making the derivations similar to the ones preformed in~\cite{krichevsky2023unconventional} (see Supplementary Information there), we can write the effective Lagrangian $\mathcal{L}_\mathrm{eff}$ of the system in the quasi-antiferromagnetic approximation as a function of the angles $\theta$ and $\phi$ of the antiferromagnetic vector $\mathbf{L}$~\cite{davydova2019ultrafast}: 
\begin{eqnarray}
    &\mathcal{L}_\mathrm{eff} = \frac{\chi}{2}\left(\left(\frac{\dot{\phi}}{\gamma}-H\right) \sin\theta + h \cos \theta \sin \phi \right)^2 +  \nonumber\\ 
    &\quad+\frac{\chi}{2}\left(\frac{\dot{\theta}}{\gamma}-h \cos\phi\right)^2 -\frac{\dot{\phi}}{\gamma}m\cos\theta + m H \cos \theta +
    \nonumber\\ 
    &\quad\quad\quad\quad\quad\quad\quad+mh \sin \theta \sin \phi+ K \sin^2 \theta \sin^2\phi, \label{Eq L_eff}
\end{eqnarray}
where $K=K_1+K_2$ is an effective anisotropy constant, $\chi=(M_1+M_2)^2/(4\Lambda M_1 M_2)\approx\Lambda^{-1}$ is the constant describing the exchange interaction between the sublattices and $\gamma$ is the gyromagnetic ratio. For a definiteness, in numerical calculations the following parameters are used: $K=300~\mathrm{erg/cm^3}$, $\chi=10^{-3}$ and $m=2~\mathrm{emu/cm^3}$ that correspond to the sample with a composition  $\mathrm{(YBiLu)_3(FeGa)_5O_{12}}$ at the room temperature.

Passing from the Lagrangian to the Hamiltonian function, we can obtain a relation for the potential energy of the system in the quasi-antiferromagnetic approximation:
\begin{eqnarray}
    &U_\mathrm{eff} = -\frac{\chi}{2}\left(h \cos\theta \sin\phi - H \sin \theta \right)^2 -\frac{\chi}{2}\left(h \cos\phi\right)^2 - \nonumber\\ 
    &-m H \cos \theta - m h \sin \theta \sin \phi- K \sin^2 \theta \sin^2\phi. \label{Eq U_eff}
\end{eqnarray}

The equilibrium state of a magnetic system is determined by the angles $\theta,\phi$ that provide the minimum of the potential energy $U_\mathrm{eff}(\theta,\phi)$ . It allows to find a magnetic phase diagram of a sample describing angle $\theta_0$ versus the in-plane magnetic field $H$ and the difference in magnetic sublatticies $m$. 

\begin{figure*}[ht]
\centering
(a)~~~~~~~~~~~~~~~~~~~~~~~~~~~~~~~~~~~~~~~~~~~~~~~~~~(b)~~~~~~~~~~~~~~~~~~~~~~~~~~~~~~~~~~~~~~~~~~~~~~~~~~(c)\\
\includegraphics[width=0.68\columnwidth]{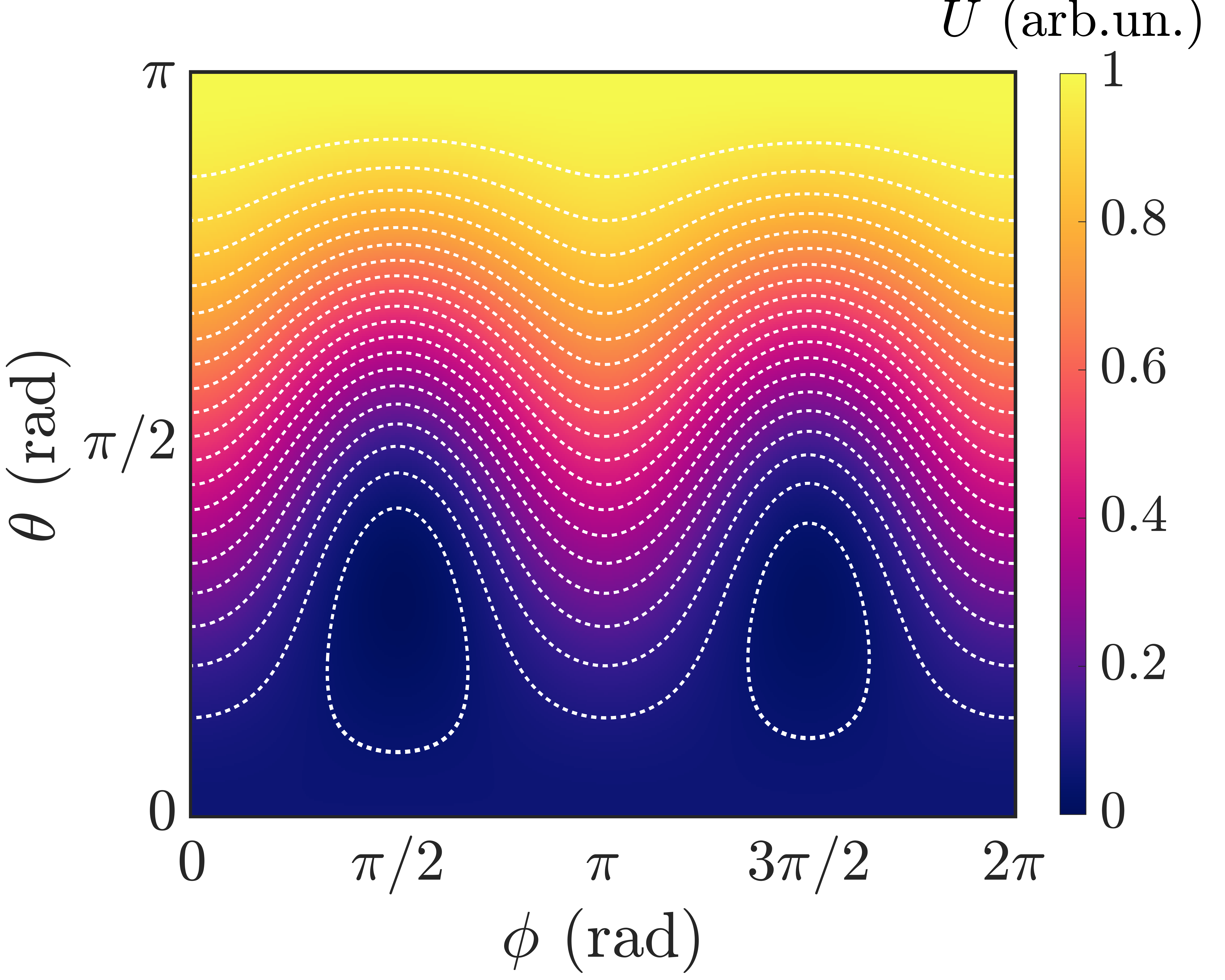}
\includegraphics[width=0.68\columnwidth]{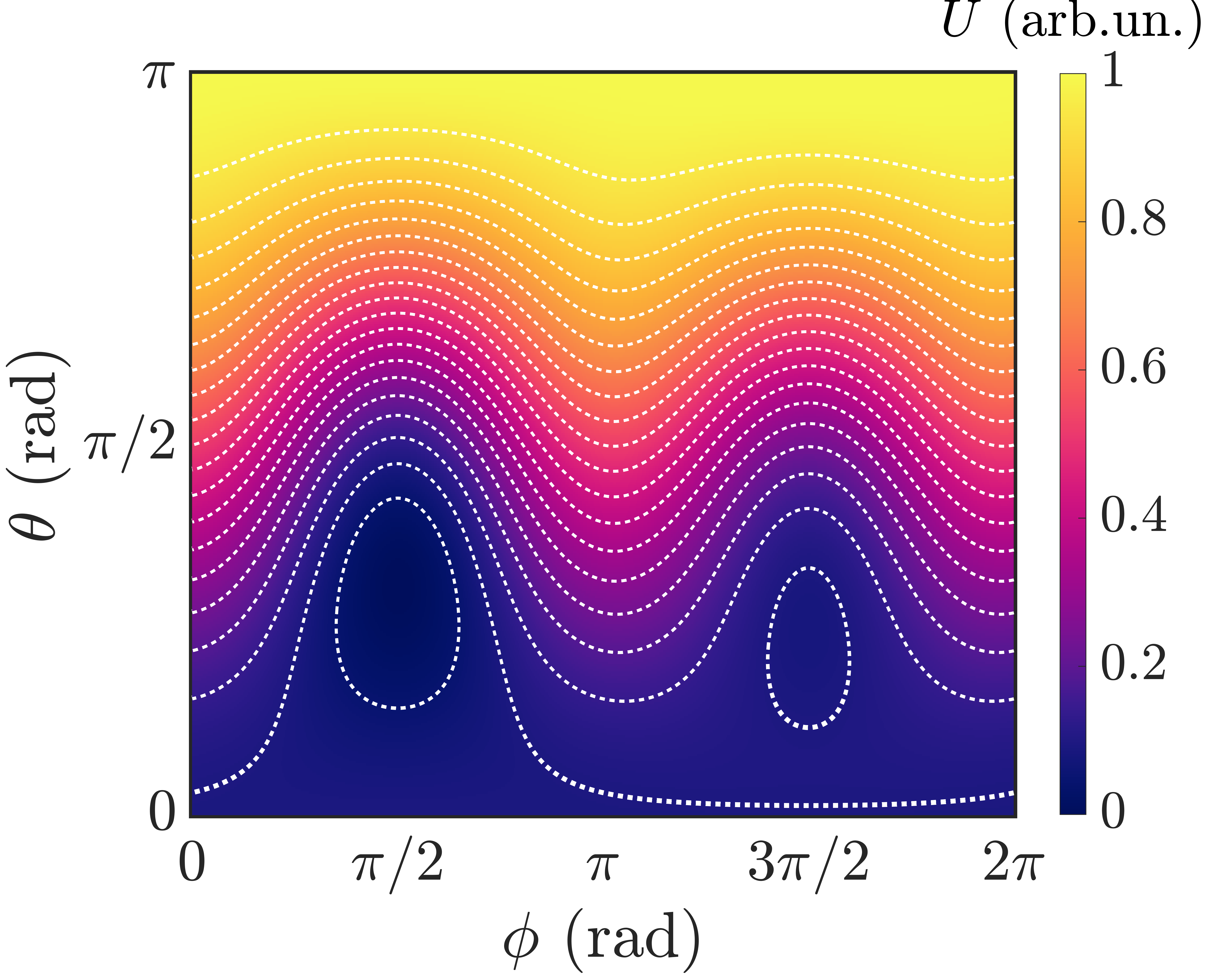}
\includegraphics[width=0.68\columnwidth]{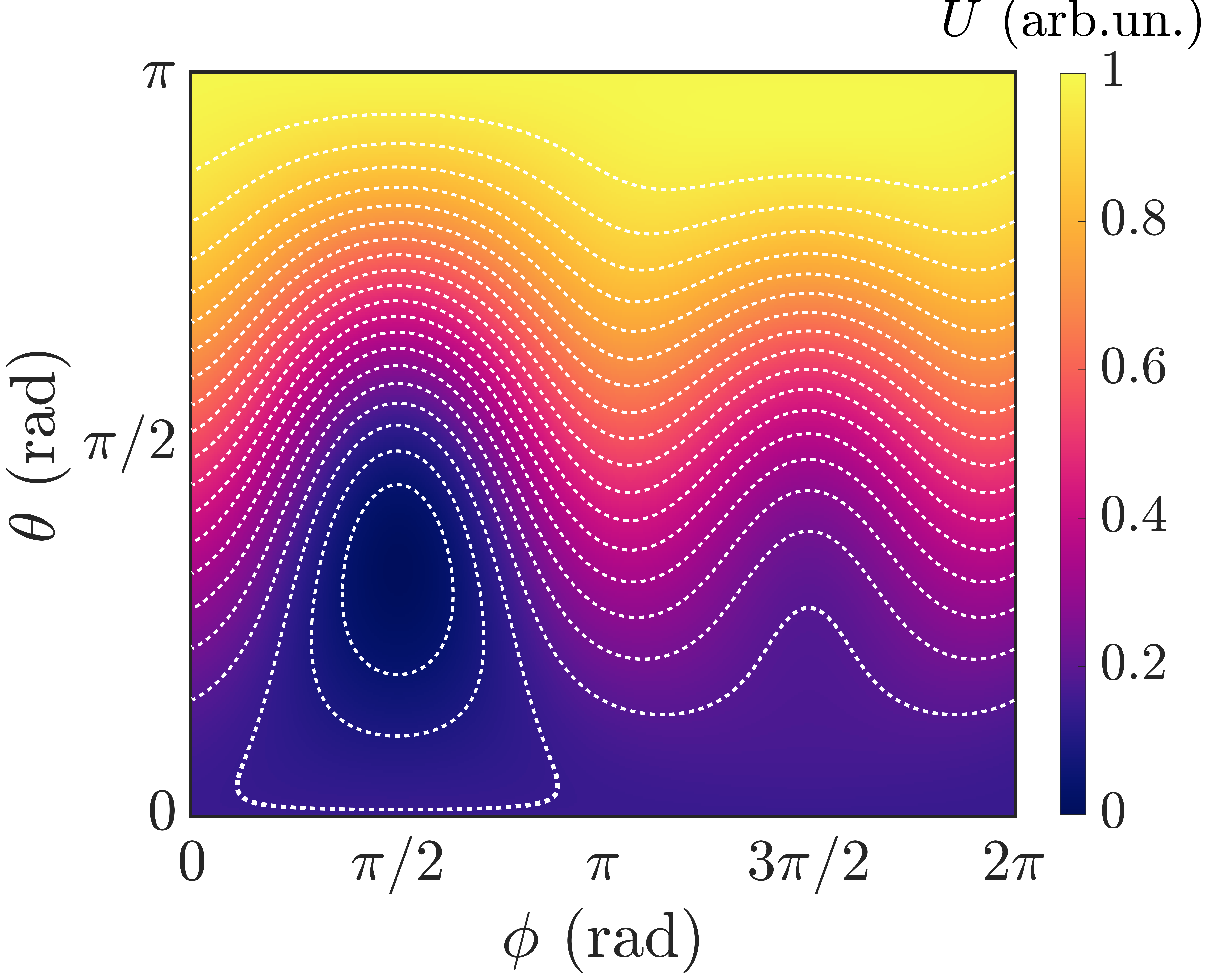}\\
\caption{False-color map of the magnetic potential energy $U_\mathrm{eff}(\theta,\phi)$ for the different values of out-of-plane $h$ component and fixed in-plane $H=2$~kOe of the external magnetic field. White lines depict the iso-levels of $U_\mathrm{eff}$. (a) $h=0$ which provides two symmetric minima of $U_\mathrm{eff}$. (b) Small $h=0.25$~kOe component which provides the asymmetry between the minima of $U_\mathrm{eff}$. (c) Relatively large $h=0.6$~kOe component providing only one minima of $U_\mathrm{eff}$. The magnetic parameters of the sample used for calculations are provided in the text.}
\label{Fig: Ueff from teta phi}
\end{figure*}

For the in-plane magnetic field ($h=0$) the magnetic phase diagram has a well-known form (see~\cite{krichevsky2023unconventional} for example) and contains two regions corresponding to the collinear and non-collinear phases (Fig.~\ref{Fig: Scheme and PD for h=0}c). The collinear phase exists in the range of $|mH| \ge 2K + \chi H^2$ and is characterized by $\theta_0=0$ for $m>0$ or $\theta_0=\pi$ for $m<0$. The non-collinear phase appears in the vicinity of a compensation point if $|mH| < 2K + \chi H^2$. In the non-collinear phase $U_\mathrm{eff}(\theta, \phi)$ is symmetric for $\phi<\pi$  and $\phi>\pi$ (Fig.~\ref{Fig: Ueff from teta phi}a) and provides the two minima with the same angle $\theta_0$ 
\begin{equation}
    \theta_0=\arccos \left( \frac{m H}{2K + \chi H^2} \right), \label{Eq teta_noncoll h0}
\end{equation}
but different angles $\phi_0=\pi/2,~3\pi/2$ Fig.~\ref{Fig: Scheme and PD for h=0}b), so that the system is bistable. 

The whole picture changes if out-of-plane magnetic field $h$ appears in the system in addition to the in-plane field $H$. These changes in terms of the Neel vector orientation are qualitatevely discussed in the next section. To understand the origins and the character of these changes, a detailed theoretical and numerical analysis is provided in next Sections in a following way. First, we fix the magnetization and the in-plane magnetic field and analyze how $U_\mathrm{eff}(\theta, \phi)$ modifies in the presence of $h$. After that, we analyze how the equilibrium $\mathbf{L}$ positions and the system stability changes due to application of $h$. That allows us finally to compare the phase diagrams of the ferrimagnets placed in the fully in-plane and two-component magnetic fields.

\section{Neel vector trajectory}

In the strictly in-plane external magnetic field the ferrimagnet has two degenerate equilibrium positions that are characterized by the same $\theta_0$ and the opposite $\phi_0=\pi/2$ and $\phi_0=3\pi/2$ values (Eq.~\eqref{Eq teta_noncoll h0}). If one parameters of the system, for example, an external magnetic field is fixed ($H=2$~kOe in Fig.~\ref{Fig: L arrows}a), and the other varies ($-6~\mathrm{emu/cm^3}<m<6~\mathrm{emu/cm^3}$), the Neel vector $\mathbf{L}$ gradually changes its orientation from $\theta_0=\pi$ to $\theta_0=0$. This can be realized in two ways, by passing through the states with $\phi_0=\pi/2$ (pink arrows in Fig.~\ref{Fig: L arrows}a), or passing through the states with $\phi_0=3\pi/2$ (violet arrows in Fig.~\ref{Fig: L arrows}a). Both trajectories are equivalent from the point of view of the energy of the states. In practice, one of paths ('up' through the states with $\phi_0=\pi/2$ or 'down' through $\phi_0=3\pi/2$ states) is selected stochastically depending on the different small deviations from the considered ideal case, such as small tilts of the external magnetic field from strictly in-plane orientation, presence of residual magnetization, cubic anisotropy etc. Similar situation is observed for a fixed magnetization $m=2~\mathrm{emu/cm^3}$ and varying external magnetic field ($-6~\mathrm{kOe}<H<6~\mathrm{kOe}$ (Fig.~\ref{Fig: L arrows}d)).

Application of the out-of-plane magnetic field breaks this degeneracy between $\phi_0=\pi/2$ and $\phi_0=3\pi/2$ states (Fig.~\ref{Fig: L arrows}b,c,e,f). As it will be shown further, in a bistability region (white color in Fig.~\ref{Fig: L arrows}) one of the states becomes more energetically favorable, and can be called stable (such states are shown by darker colors in Fig.~\ref{Fig: L arrows}b,c,e,f) that the other one (shown by light colors), which may be treated as a quasi-stable state. Under the magnetization change from $m=-6~\mathrm{emu/cm^3}$ to $m=6~\mathrm{emu/cm^3}$ for a fixed inclined magnetic field (Fig.~\ref{Fig: L arrows}b,c) first $\mathbf{L}$ gradually changes according to the more energetically favorable trajectory (dark violet arrows). Passing through the $m=0$ point swaps the stability of the 'up' and 'down' positions, so $\mathbf{L}$ continues its path through a less favourable quasi-stable states (light violet arrows) unless it reaches the bistability border ($m\approx 2.5~\mathrm{emu/cm^3})$, grey region) where these states vanish. At this point, $\mathbf{L}$ flips to the opposite orientation and continues its path through the stable states (dark pink arrows). If one changes the magnetization in the opposite way, from $m=6~\mathrm{emu/cm^3}$ to $m=-6~\mathrm{emu/cm^3}$, $\mathbf{L}$ again passes through the more favourable stable states (dark pink arrows in Fig.~\ref{Fig: L arrows}b,c) which transform to a quasi-stable less favourable ones (light pink  arrows in Fig.~\ref{Fig: L arrows}b,c) at the compensation point $m=0$. Finally, $\mathbf{L}$ reaches the other bistability boarder ($m\approx -2.5~\mathrm{emu/cm^3}$, grey region) where it flips to the opposite direction corresponding to the stable states. 

Thus, a continuous change of the magnetization across the compensation point results in the hysteresis behaviour of the Neel vector orientation. Similar  hysteresis behaviour of the Neel vector $\mathbf{L}$ is observed in the presence of the inclined magnetic field with both in-plane and out-of-plane components simultaneously changed (Fig.~\ref{Fig: L arrows}c,f). On the contrary, if out-of-plane magnetic field is fixed, and in-plane field varies (Fig.~\ref{Fig: L arrows}e), the states $\phi_0=\pi/2$ are preferable.

\begin{figure*}[ht]
\centering
(a)~~~~~~~~~~~~~~~~~~~~~~~~~~~~~~~~~~~~~~~~~(b)~~~~~~~~~~~~~~~~~~~~~~~~~~~~~~~~~~~~~~~~~~~~~~~~~~~~~~~~~~~~~~~~~~~~(c)\\
\includegraphics[height=4cm]{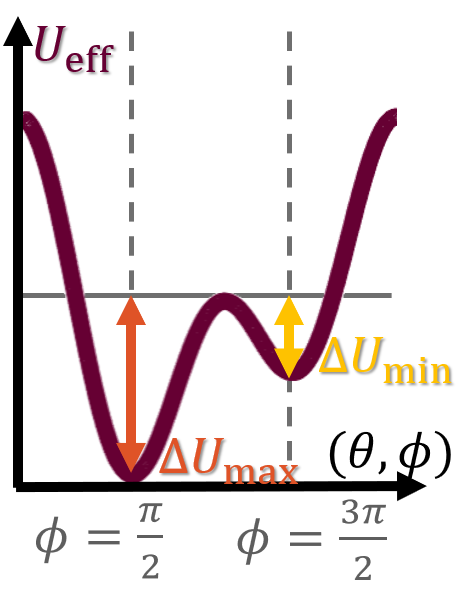}
\includegraphics[width=0.41\linewidth]{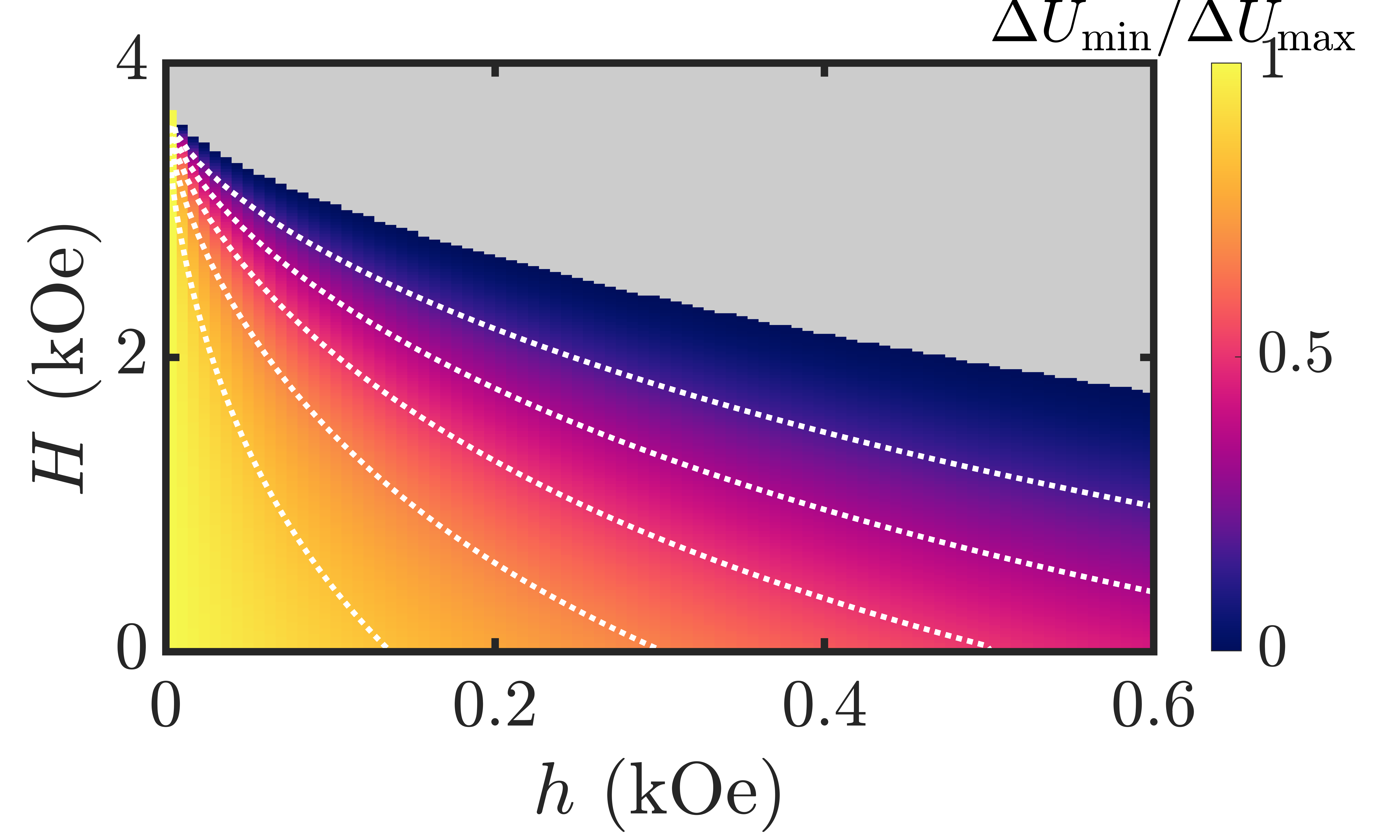}
\includegraphics[width=0.41\linewidth]{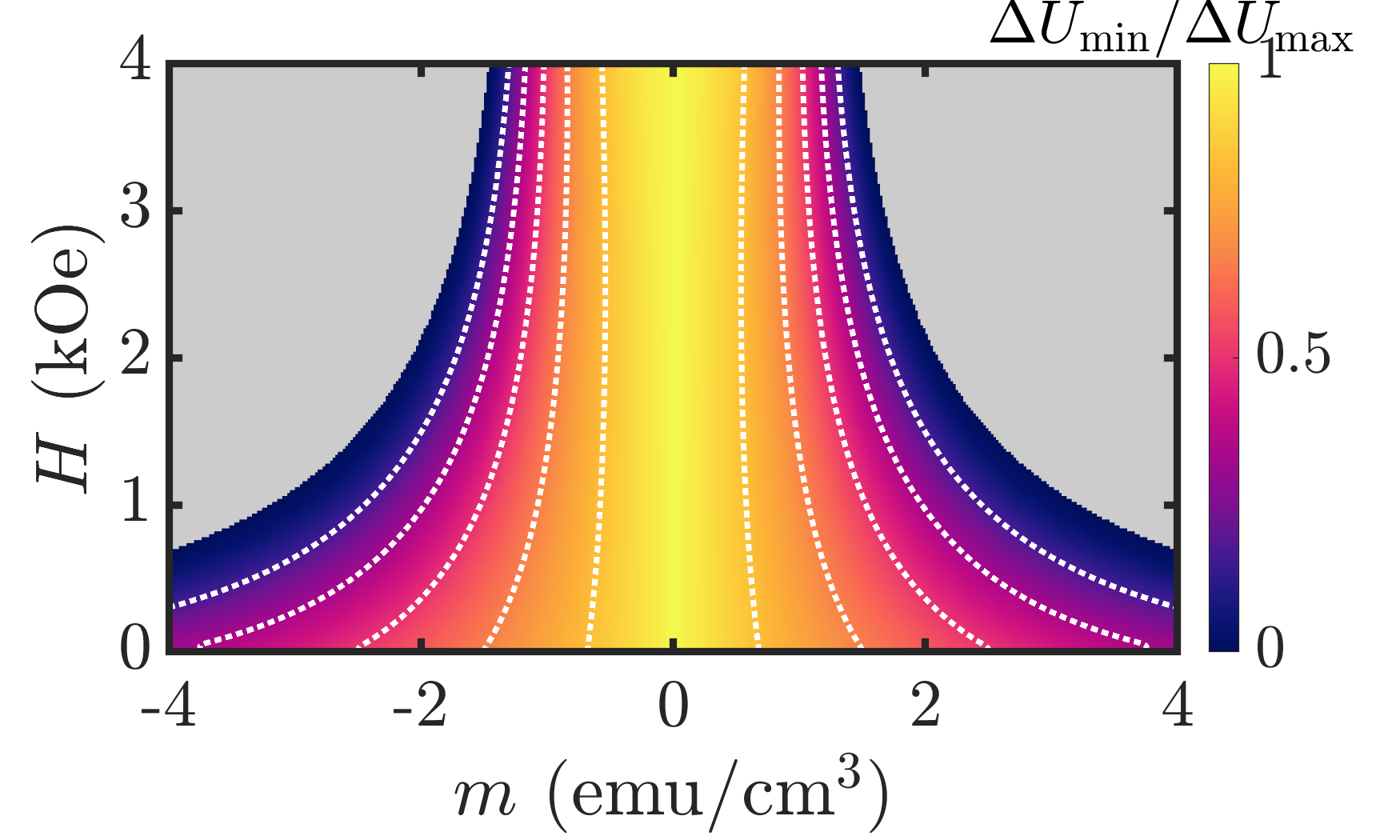}
\caption{The relative value of a potential barrier $\Delta U_\mathrm{min}/\Delta U_\mathrm{max}$ between the two equilibrium states ($\phi=\pi/2$ and $\phi=3\pi/2$). (a) Schematic depiction of the barrier change under application of out-of-plane field. (b) Dependence of $\Delta U_\mathrm{min}/\Delta U_\mathrm{max}$ on $h$ and $H$ fields for a fixed $m=2~\mathrm{emu/cm^3}$ and (b) dependence of $\Delta U_\mathrm{min}/\Delta U_\mathrm{max}$ on $H$ and $m$ for a fixed $h=0.4$~kOe. White dashed curves show the isolines with 0.2 step. Gray area denote the region with a single potential energy minimum.}
\label{Fig: dU/U}
\end{figure*}

\section{Effective energy of a ferrimagnetic in the incline magnetic field}

For $h=0$ the function $U_\mathrm{eff}(\theta, \phi)$ in the non-collinear state has two symmetric minima as it was mentioned above (Fig.~\ref{Fig: Ueff from teta phi}a). If an out-of-plane field $h$ is applied, the two minima still exist located at $\phi_0=\pi/2,~3\pi/2$, but different $\theta_0$ values. An important change in $U_\mathrm{eff}(\theta, \phi)$ function is that 
one of the minima of $U_\mathrm{eff}(\theta, \phi)$ becomes more pronounced, and the other gets swallow (Fig.~\ref{Fig: Ueff from teta phi}b). This means that the state with an out-of-plane component of the net magnetization $\mathbf{M}=\mathbf{M_1}+\mathbf{M_2}$ directed along $\mathbf{h}$ field ($(\mathbf{M,h})>0$) is preferable. However, the second equilibrium state, with out-of-plane component of $\mathbf{M}$ directed oppositely to $\mathbf{h}$ ($(\mathbf{M,h})<0$) still exists. Thus, the bistability remains. With the increase of $h$, the asymmetry grows and finally the second minima of $U_\mathrm{eff}(\theta, \phi)$ disappears (Fig.~\ref{Fig: Ueff from teta phi}c) and the system becomes monostable with only one stable state at $\phi_0=\pi/2$ and ($(\mathbf{M,h})>0$).

Thus, application of the out-of-plane magnetic field diminishes the depth of a potential barrier corresponding to the oppositely directed magnetization (compare Fig.~\ref{Fig: Ueff from teta phi}a,b and the two minima in Fig.~\ref{Fig: Ueff from teta phi}b). While for the quite small magnetic fields $h$ the second minimum of the effective potential energy $U_\mathrm{eff}$ still exists mathematically (Fig.~\ref{Fig: Ueff from teta phi}b), its depth becomes extremely small. Figure~\ref{Fig: dU/U} illustrates the differences of the potential barriers between the two equilibrium states $U_{\phi=\pi/2}$ and $U_{\phi=3\pi/2}$, which are determined as minima of $U_\mathrm{eff}$ (Eq.~\eqref{Eq U_eff}). $\Delta U_\mathrm{min}$ characterizes the minimal energy required to overcome a potential barrier and to switch from the higher-energy to the low-energy state . In other words, $\Delta U_\mathrm{min}$ is the depth of the most swallow minimum (see a sketch in Fig.~\ref{Fig: dU/U}a). For a convenience, it is normalized on the depth of the other minima $\Delta U_\mathrm{max}$. Both $\Delta U$ are calculated for the optimal trajectory between the $U_{\phi=\pi/2}$ and $U_{\phi=3\pi/2}$ states.

In the absence of the out-of-plane magnetic field $h=0$ (Fig.~\ref{Fig: dU/U}b) or at the magnetization compensation point $m=0$ (Fig.~\ref{Fig: dU/U}c) the two states with $\phi=\pi/2$ and $\phi=3\pi/2$ are equivalent to each other and have the same energy (Fig.~\ref{Fig: Ueff from teta phi}). Application of the out-of-plane magnetic field $h$ makes one of the minima more swallow, and the difference between the minima depth grows. Figure~\ref{Fig: dU/U}b,c illustrates that the potential barrier between the two states is less than $20\%$ of its depth in a rather wide $H$ range in the vicinity of the transition between the bistable and monostable states. For example, quite moderate out-of-plane magnetic field $h\sim40$~mT makes $\Delta U_\mathrm{min}/\Delta U_\mathrm{max}<20\%$ in the region $1.5~\mathrm{kOe}<H<2~{kOe}$ for a fixed $m=2~\mathrm{emu/cm^3}$ (Fig.~\ref{Fig: dU/U}b,c). This region grows with the increase of $h$ ($1~\mathrm{kOe}<H<1.8~{kOe}$ for $h=0.6$~kOe, $m=2~\mathrm{emu/cm^3}$ Fig.~\ref{Fig: dU/U}b) or as one moves away from the compensation point ($0.25~\mathrm{kOe}<H<0.7~{kOe}$ for $h=40$~kOe, $m=4~\mathrm{emu/cm^3}$, Fig.~\ref{Fig: dU/U}c). Although the bistability in this region exists in theory, the higher-energy state would be quite unstable in practice.

\begin{figure*}[ht]
\centering
(a)~~~~~~~~~~~~~~~~~~~~~~~~~~~~~~~~~~~~~~~~~~~~~~~(b)~~~~~~~~~~~~~~~~~~~~~~~~~~~~~~~~~~~~~~~~~~~~~~~(c)\\
\includegraphics[width=0.32\linewidth]{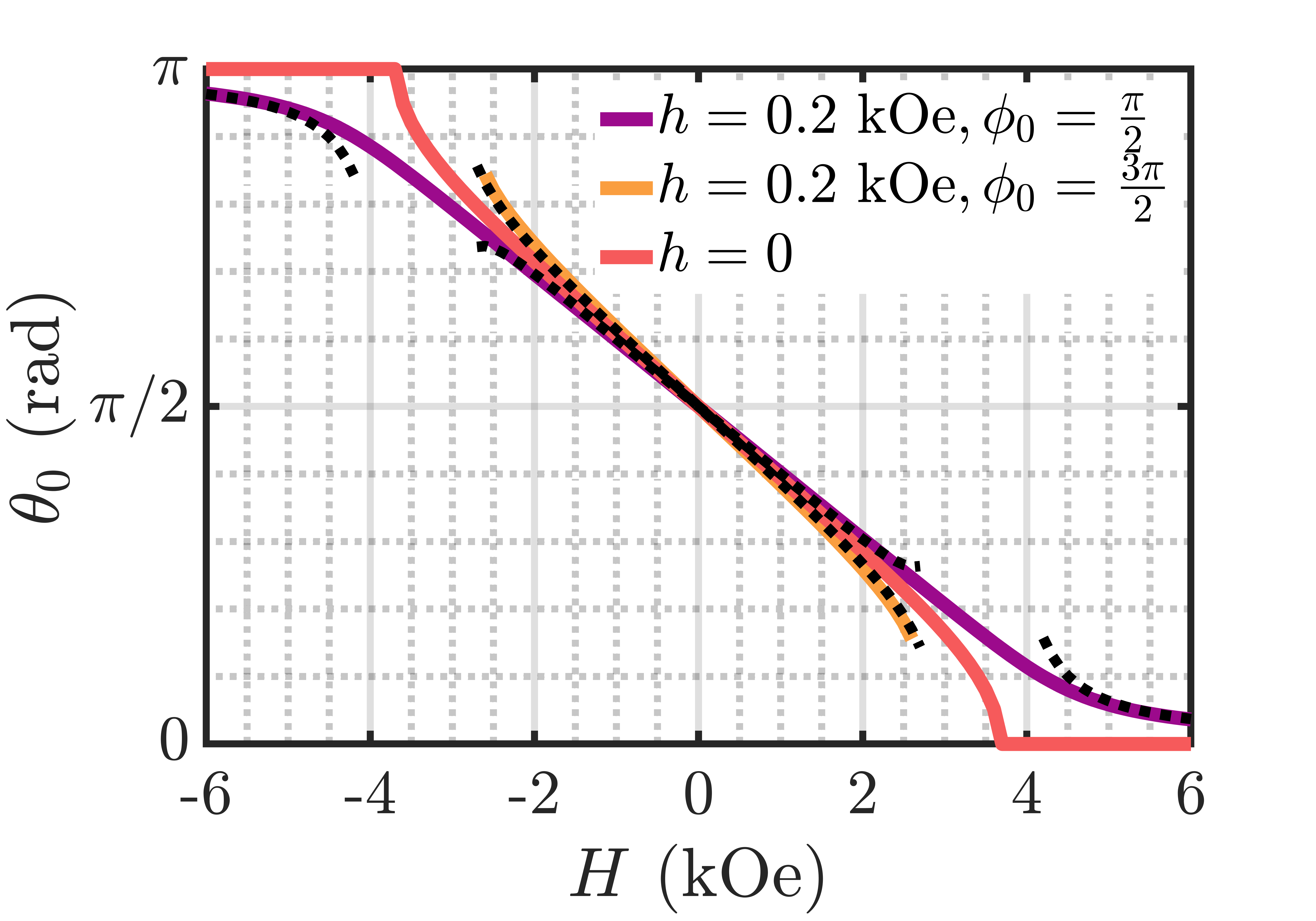}
\includegraphics[width=0.32\linewidth]{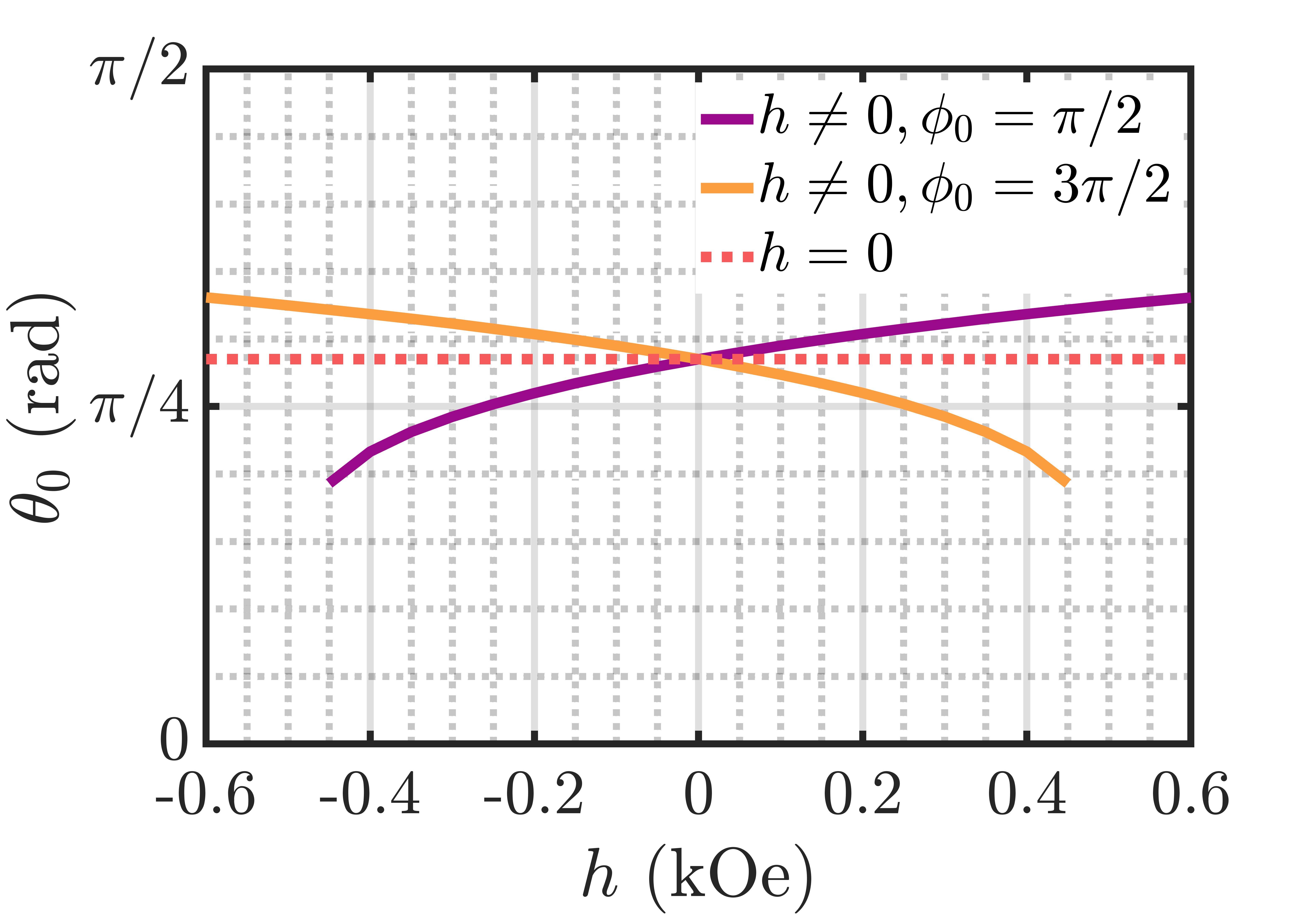}
\includegraphics[width=0.32\linewidth]{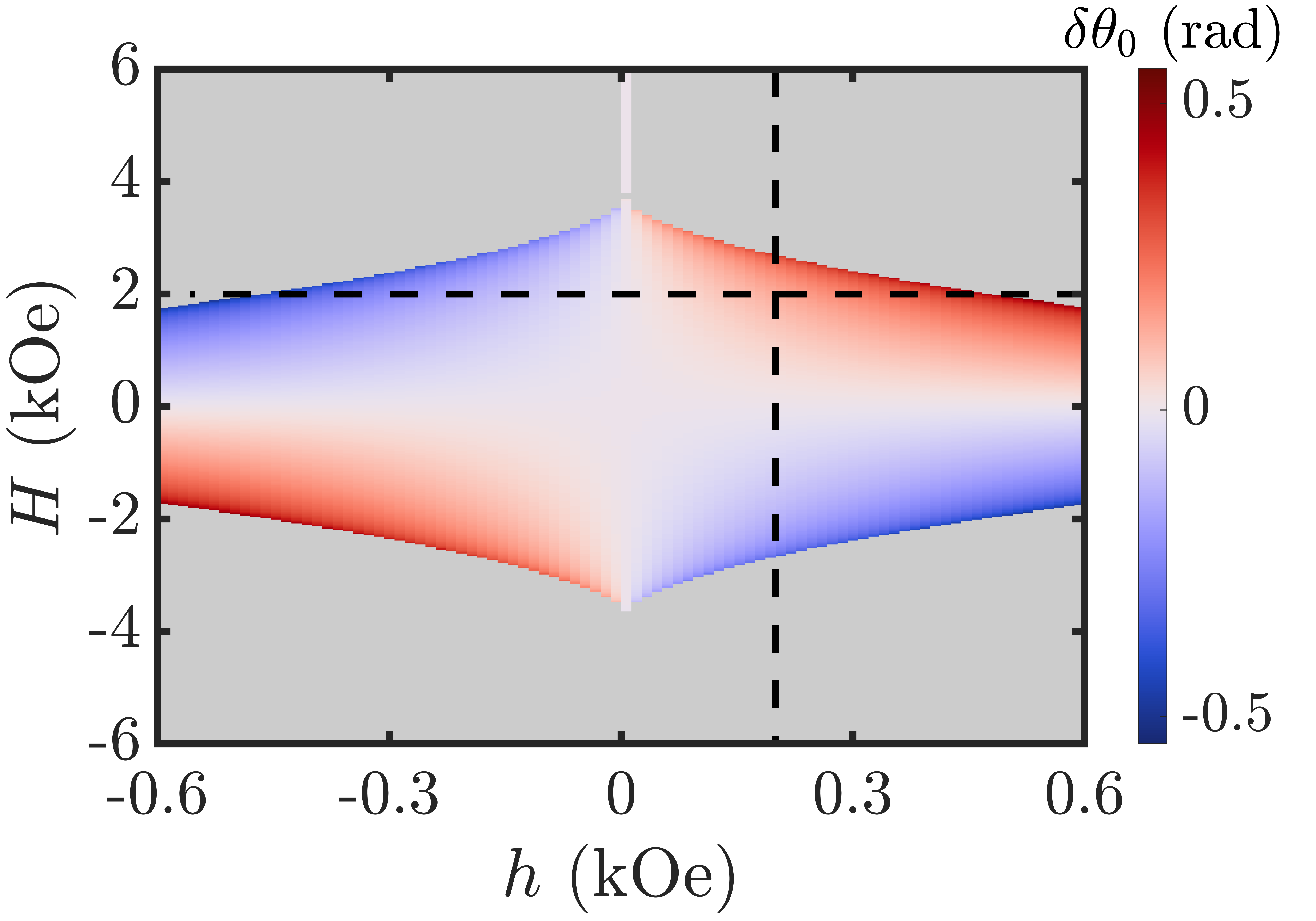}\\
(d)~~~~~~~~~~~~~~~~~~~~~~~~~~~~~~~~~~~~~~~~~~~~~~~~~~~~~~~~~~~~~~~~~~~~~~~~~~~~~~~~~~~~~~~(e)\\
\includegraphics[width=0.8\columnwidth]{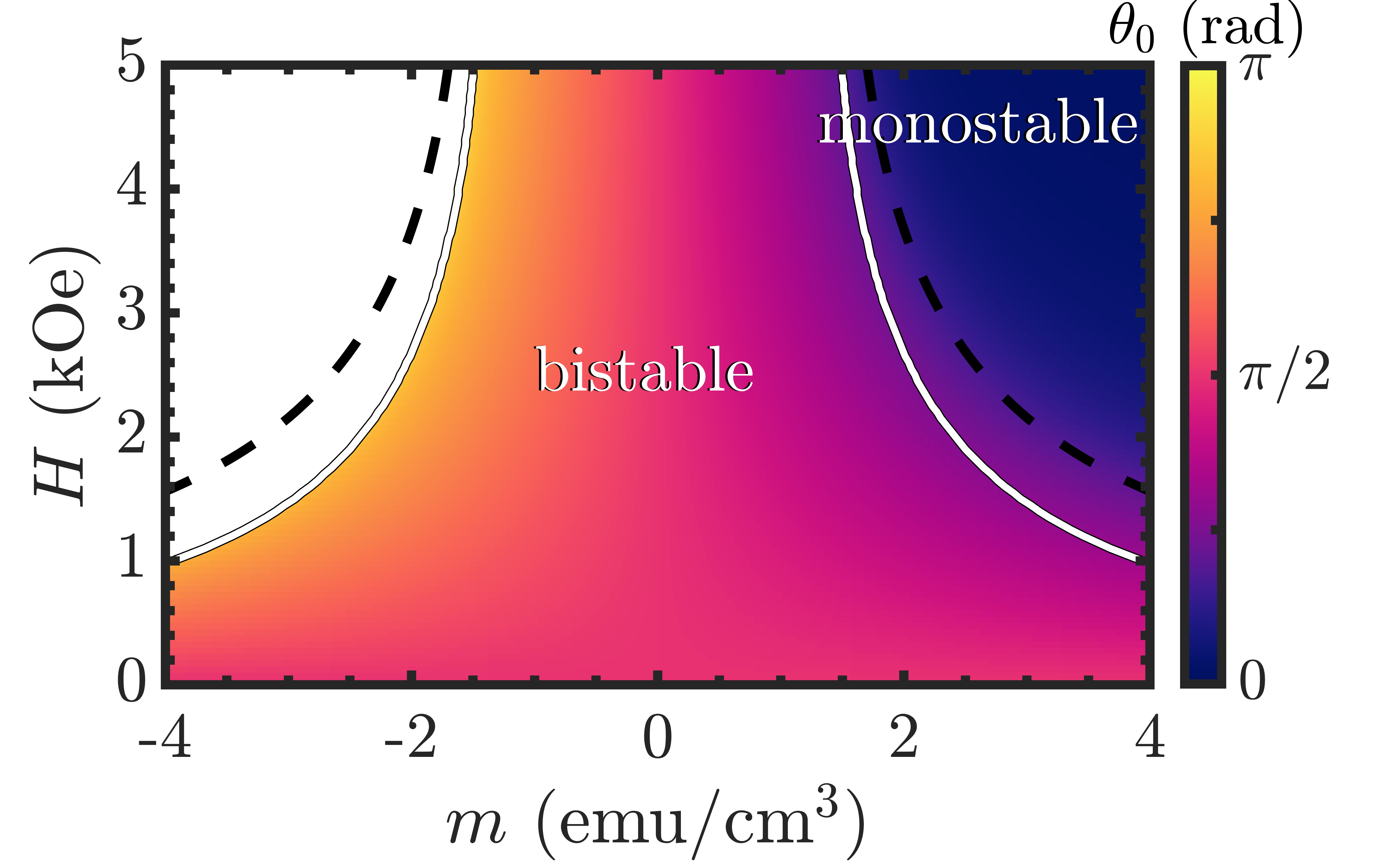}~~~
\includegraphics[width=0.8\columnwidth]{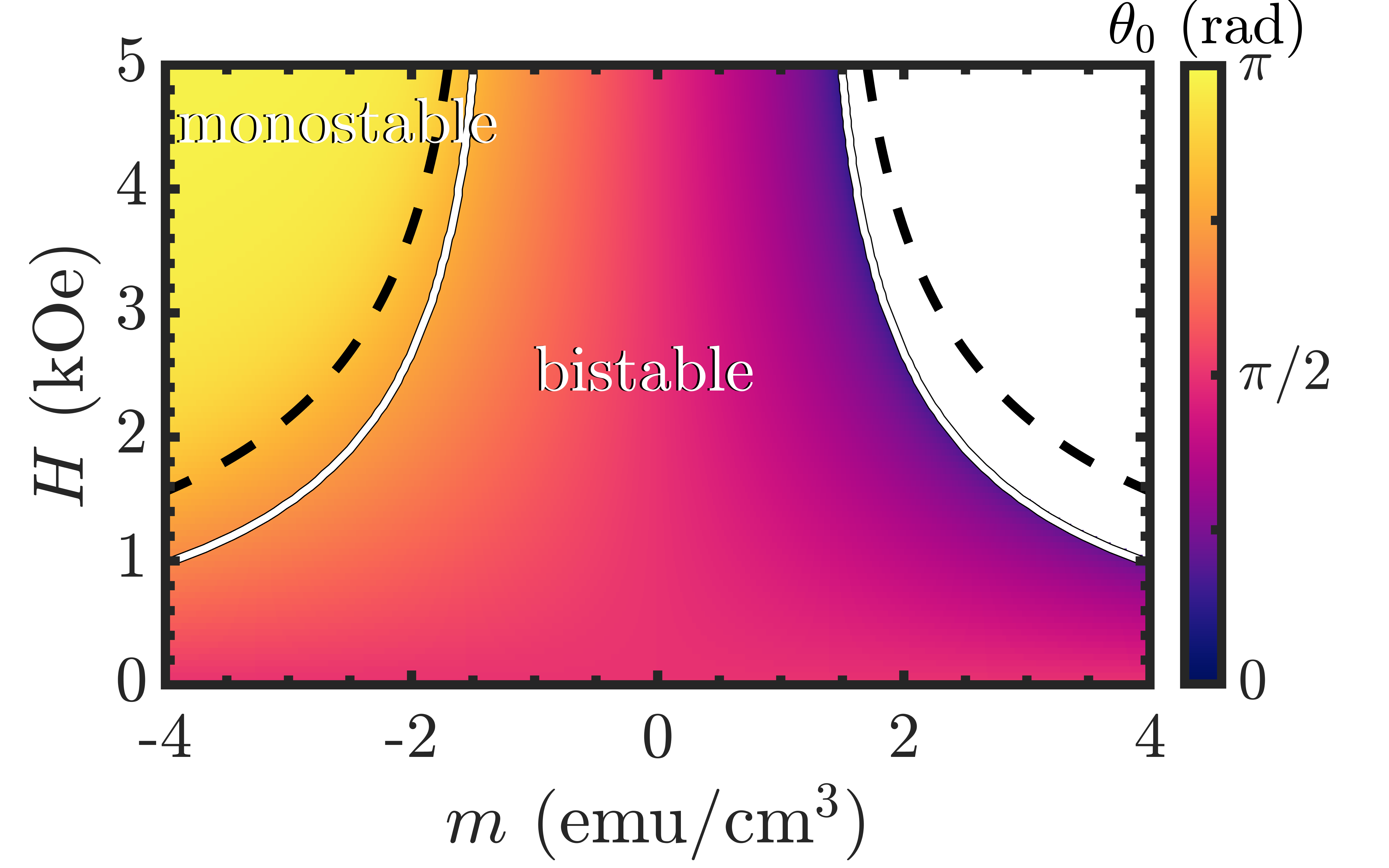}\\
\caption{Equilibrium positions $\theta_{0,~\phi=\pi/2,3\pi/2}$ in the presence of the out-of-plane magnetic field $h$. (a) Values of $\theta_{\phi=\pi/2,3\pi/2}(H)$ for non-zero $h=0.2$~kOe (solid orange and violet lines). Pink line shows $\theta_0(H)$ for $h=0$ that is the same for $\phi_0=\pi/2, 3\pi/2$. Dotted black lines show the approximations Eqs.~\eqref{Eq delta_teta approx small h},~\eqref{Eq delta_teta quasi-collinear}. (b) Tuning the number and position of equilibrium states $\theta_{0\phi=\pi/2,3\pi/2}$ by applying the out-of-plane $h$ field (solid orange and violet lines) in $m=2~\mathrm{emu/cm^3}$ and $H=2$~kOe configuration. Pink dashed line shows the level of $\theta_0$ for $h=0$. (c) The difference between the equilibrium $\theta_0$ values $\delta \theta_0 = \theta_{0(\pi/2)}-\theta_{0(3\pi/2)}$ vs. in-plane $H$ and out-of-plane $h$ magnetic field components. Grey area shows the region where only one solution exists. Dashed black lines correspond to the cross-sections shown in (a) and (b) plots. (d) $\theta_{0,~\phi=\pi/2}$ and (e) $\theta_{0,~\phi=3\pi/2}$ in $(m,H)$ coordinates for a fixed out-of-plane magnetic field $h=0.2$~kOe. White lines show the boarders between the bistable and monostable regimes. Dashed black lines show the same boarder coinciding with a phase transition obtained for $h=0$ (see Fig.~\ref{Fig: Scheme and PD for h=0}).}
\label{Fig: teta from H}
\end{figure*}

\section{Equilibrium states of a ferrimagnet in the incline magnetic field}

As the presence of the small out-of-plane magnetic field $h$ can change the number and position of the magnetization equilibrium states, it is important to analyze its impact in more detail. The minima of Eq.~\eqref{Eq U_eff} are determined by the following transcendental equation:
\begin{eqnarray}
    -(K+\frac{\chi}{2} (H^2-h^2)) \sin 2\theta_0 + mH \sin \theta_0 \nonumber\\
    \pm \chi Hh \cos 2\theta_0 \mp mh \cos \theta_0=0, \label{Eq eq_for_teta}
\end{eqnarray}
where $\pm$ signs correspond to the two minima at $\phi_0=\pi/2,~3\pi/2$, respectively. Numerical solution of this equation for the fixed $H$ or $h$ values is shown in Fig.~\ref{Fig: teta from H}.

Making the expansion of Eq.~\eqref{Eq eq_for_teta} near the equilibrium $\theta_0$  determined by Eq.~\eqref{Eq teta_noncoll h0}  allows one to obtain a shift of the equilibrium angle $\Delta \theta_0$ which is proportional to $h$:
\begin{equation}
    \Delta \theta_0 = \pm h \frac{m \cos \theta_0+\chi H \cos 2\theta_0}{(\sin^2 \theta_0 - \frac{\chi h^2}{2 K + \chi H^2} \cos 2\theta_0)( 2 K + \chi H^2)}, \label{Eq delta_teta approx}
\end{equation}
where $\pm$ signs correspond to the two solutions with $\phi_0=\pi/2,~3\pi/2$, respectively. For rather small values of $h^2 \ll 2K / \chi$ one may simplify Eq.~\eqref{Eq delta_teta approx} to:
\begin{equation}
    \Delta \theta_0 = \pm \frac{h}{\sin^2 \theta_0} \frac{m \cos \theta_0-\chi H \cos 2\theta_0}{2 K + \chi H^2}. \label{Eq delta_teta approx small h}
\end{equation}
These $\theta_0 \pm \Delta \theta_0$ values describe analytically the bistable regime (see the central part of Fig.~\ref{Fig: teta from H}a where the violet and orange curves coexist).

Equation~\eqref{Eq delta_teta approx small h} shows that the shift of the equilibrium position $\Delta \theta_0$ significantly grows while $\mathbf{L}$ vector tends to the film plane ($xz$ plane) since $\sin \theta_0$ becomes close to zero. However, Eq.~\eqref{Eq delta_teta approx} was obtained under the assumption of smallness of $\Delta \theta_0$ and is not valid for the simultaneously small $\sin \theta_0 \approx 0$ and large $\Delta \theta_0$. Thus, this region of phase diagram with the close to the in-plane orientation of $\mathbf{L}$ vector should be analyzed in a different way. Assuming that $\theta_0 \approx 0~\mathrm{or}~\pi$ one may simplify Eq.~\eqref{Eq delta_teta approx} and obtain that the bistablity vanishes. Only one equilibrium state for these parameters exist:
\begin{eqnarray}
    \theta_0 = \left\lvert h \frac{m -\chi H }{2 K + \chi (H^2-h^2) - mH}\right\rvert, ~mH>0, \nonumber \\
    \label{Eq delta_teta quasi-collinear} \\
    \theta_0 = \pi -\left\lvert h \frac{m +\chi H }{2 K + \chi (H^2-h^2) + mH}\right\rvert, ~mH<0, \nonumber
\end{eqnarray}
and $\phi_0 = \pi/2$ or $3\pi/2$ for $h>0$ and $h<0$, correspondingly. This monostable state is realized in the region of phase diagram that correspond to a collinear state in the absence of $h$ (see the pink curve and $\theta_0=0, \pi$ region).

\begin{figure*}[ht]
\centering
(a)~~~~~~~~~~~~~~~~~~~~~~~~~~~~~~~~~~~~~~~~~~~~~~~~~~~~~~~~~~~~~~~~~~~~~~~~~~~~~~~~~~~~~~~(b)\\
\includegraphics[width=0.49\linewidth]{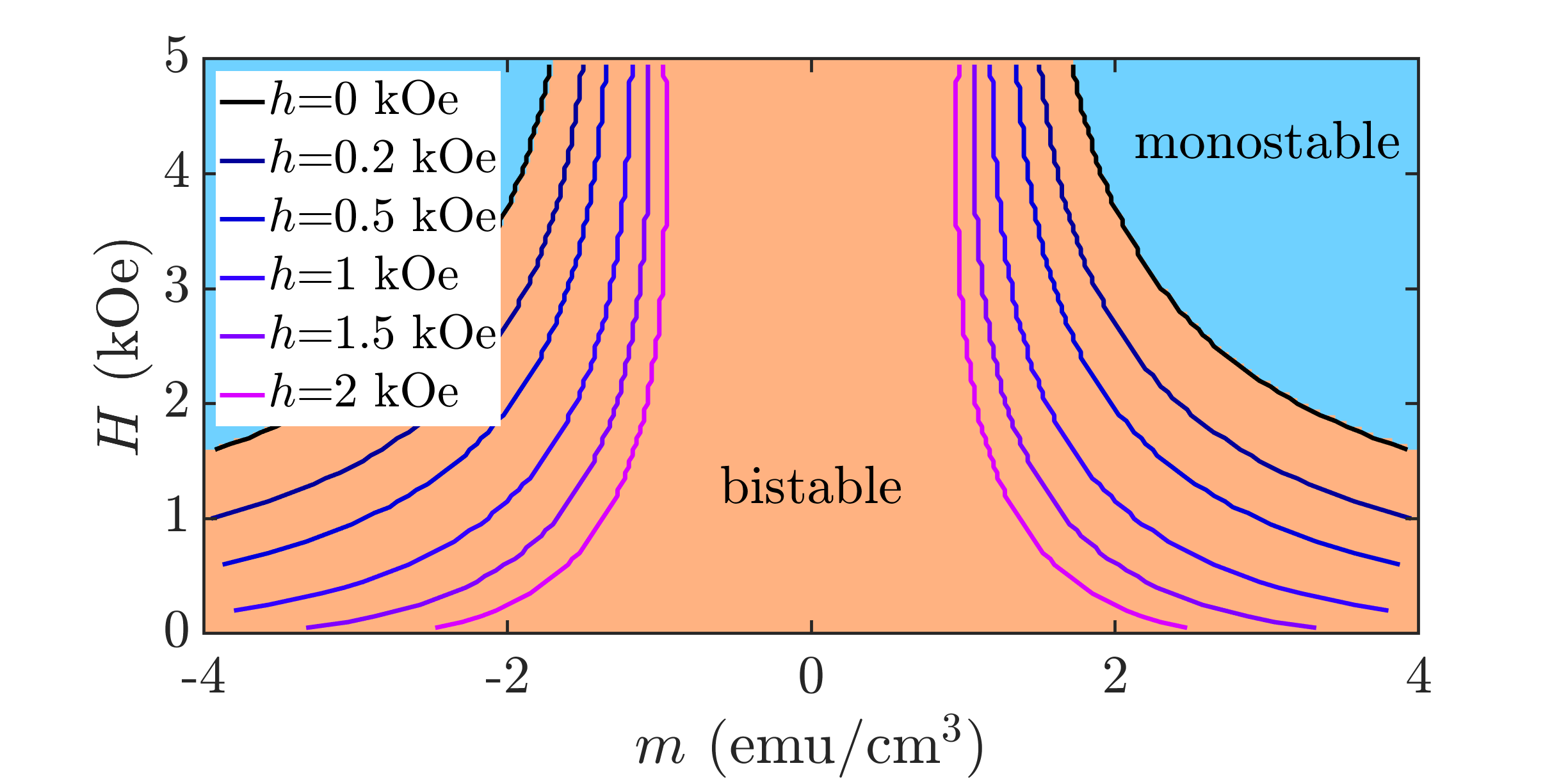}
\includegraphics[width=0.49\linewidth]{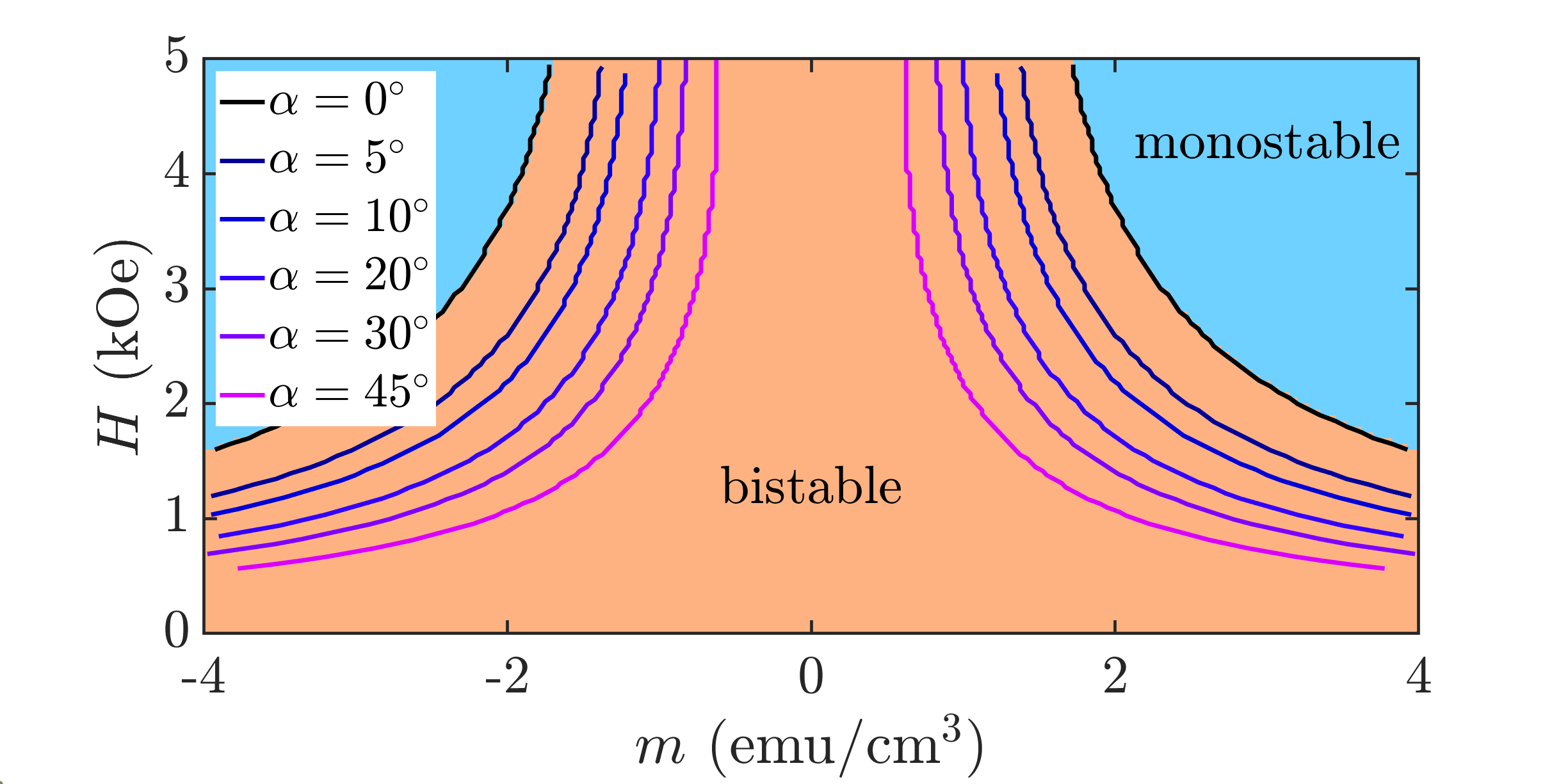}
\caption{Phase diagram of a ferrimagnet in a vicinity of a compensation point. Blue region denotes collinear phase and pink region denotes the non-collinear phase for the case $h=0$. Solid lines (see the legends) show the shift of the bistability region boarder under application of different out-of-plane magnetic fields $h$ (a) via independent magnet in addition to the in-plane field $H$ (b) via the sample tilt at an angle $\alpha$.}
\label{Fig: phase diagram}
\end{figure*}

Figure~\ref{Fig: teta from H} demonstrates bistable and mono-stable equilibrium positions of the Neel vector $\mathrm{L}$ for different values of the two-component external magnetic field. There is a good agreement between the numerically calculated through Eq.~\eqref{Eq U_eff} (orange and violet curves in Fig.~\ref{Fig: teta from H}a) and the approximations by~Eqs.~\eqref{Eq delta_teta approx small h},~\eqref{Eq delta_teta quasi-collinear} (black dashed curves in Fig.~\ref{Fig: teta from H}a) of the equilibrium angles $\theta_{0(\phi=\pi/2,3\pi/2)}$. 

For the strictly in-plane magnetic field (pink curve in Fig.~\ref{Fig: teta from H}) $\theta_0$ values for the minima determined by $\phi_0=\pi/2,~3\pi/2$ coincide with each other, and the phase transitions between the collinear and non-collinear states are clearly seen as the kinks of $\theta_0(H)$ curve. The presence of out-of-plane $h$ field modifies $\theta_0(H)$ dependence significantly (compare violet and pink curves in Fig.~\ref{Fig: teta from H}a). First of all, one may see that there is no kink for $\theta_0(H)$ dependence in the presence of the small magnetic field (violet curve in Fig.~\ref{Fig: teta from H}a). As such a kink is a characteristic feature of the second-kind phase transition, one may suppose that such transition vanishes due to the application of the out-of-plane magnetic field. At the same time, according to Eq.~\eqref{Eq delta_teta quasi-collinear} $\theta_0(H)\ne \arctan(h/H)$ for any finite value of $H$. This means, that the Neel vector is not collinear with the orientation of the external magnetic field even in the monostable regime. This is in a contrast with a collinear phase realized for the in-plane external magnetic field where $\mathbf{L}$ is aligned along $\mathbf{H}$. The question of collinearity of the sublattices for a bistable and monostable regimes realized for the two-component field will be discussed in details below.

The two equilibrium states with $\phi_0=\pi/2,~3\pi/2$ in the non-collinear phase of the in-plane configuration ($h=0$) are degenerate and have the same $\theta_0$ values (pink curve in Fig.~\ref{Fig: teta from H}). Under application of any values of out-of-plane magnetic field this degeneracy between the two states is lifted and $\theta_0$ differs for these states (orange and violet curves in Fig.~\ref{Fig: teta from H}a,b). Moreover, one of the solutions providing $\mathbf{M}$ aligned oppositely to $\mathbf{h}$ field (orange line in Fig.~\ref{Fig: teta from H}a) exists only in a certain range of $H$ ($-2.5<H<2.5$~kOe for the considered parameters in Fig.~\ref{Fig: teta from H}a) and disappears as $\mathbf{L}$ tilts more towards $xz$ plane under application of stronger $H$. Note that the stability loss condition, i.e. point where the second equilibrium state disappears, differs from the point of second-kind phase transition between the collinear and non-collinear phase in $h=0$ state.

As shown in Fig.~\ref{Fig: teta from H}b, for a fixed in-plane magnetic field $H=2$~kOe application of an order smaller $h$ significantly changes the equilibrium $\theta$ values by up to 0.3~rad, and also allows to switch between the bistable ($|h|<0.45$~kOe) and mono-stable ($|h|>0.45$~kOe) regimes.

Figure~\ref{Fig: teta from H}c summarizes how the difference between the equilibrium positions $\delta \theta_0 = \theta_{0(\pi/2)}-\theta_{0(3\pi/2)}$ and the boundary between the monostable and bistable regimes can be controlled by the out-of-plane $h$ fields for different values of the in-plane magnetic field $H$. By varying the out-of-plane $h$ from $h=0$ to $h=0.6$~kOe one might move the bistability boundary from $H\approx 4$~kOe to $H\approx2$~kOe, correspondingly. An ability to modify phase diagram will be analyzed further in more details. 

The direction of $\mathbf{h}$ is responsible for making $\theta_{0(\pi/2)}$ or $\theta_{0(3\pi/2)}$ state preferable. Figure~\ref{Fig: teta from H}d,e show how the $\theta(m,H)$ dependence modifies (see Fig.~\ref{Fig: Scheme and PD for h=0} for a comparison with $h=0$ case) in the presence of out-of-plane magnetic field. One might see that the state with $(\mathbf{M,h})>0$) is preferable, so that $\theta_{0(\pi/2)}$ and $\theta_{0(3\pi/2)}$ solutions remain for $m>0$ and $m<0$, correspondingly. The value $\theta_0$ gradually changes throughout the whole region where the solution exist and does not exhibit a kink. The phase transition lines that existed for $h=0$ case do not coincide with the stability loss lines (the boundary of a white region where the solution disappears, see Fig.~\ref{Fig: teta from H}d,e) that illustrates the possibility to modify the phase diagram. Let us study this phenomenon in mode detail.

\section{Phase diagram for the incline magnetic field}

For the strictly in-plane magnetic field the phase diagram in terms of $(m,H)$ coordinates is well known (Fig.~\ref{Fig: Scheme and PD for h=0}). It contains the non-collinear phase arising near the compensation point $m=0$ and the collinear phase existing for quite large values of $m$ and $H$. The boundary between these phases is described by the condition $|mH| = 2K + \chi H^2$. The non-collinear phase is characterized by the bistability of the antiferromagnetic vector position while there is only a single stable state the collinear phase (see the  pink and blue areas in Fig.~\ref{Fig: phase diagram}, correspondingly). Thus the condition $|mH| = 2K + \chi H^2$ describes the boundary between these two regimes, too (see the black line in Fig.~\ref{Fig: phase diagram}).

\begin{figure*}[ht]
\centering
(a)~~~~~~~~~~~~~~~~~~~~~~~~~~~~~~~~~~~~~~~~~~~~~~~(b)~~~~~~~~~~~~~~~~~~~~~~~~~~~~~~~~~~~~~~~~~~~~~~~(c)\\
\includegraphics[width=0.68\columnwidth]{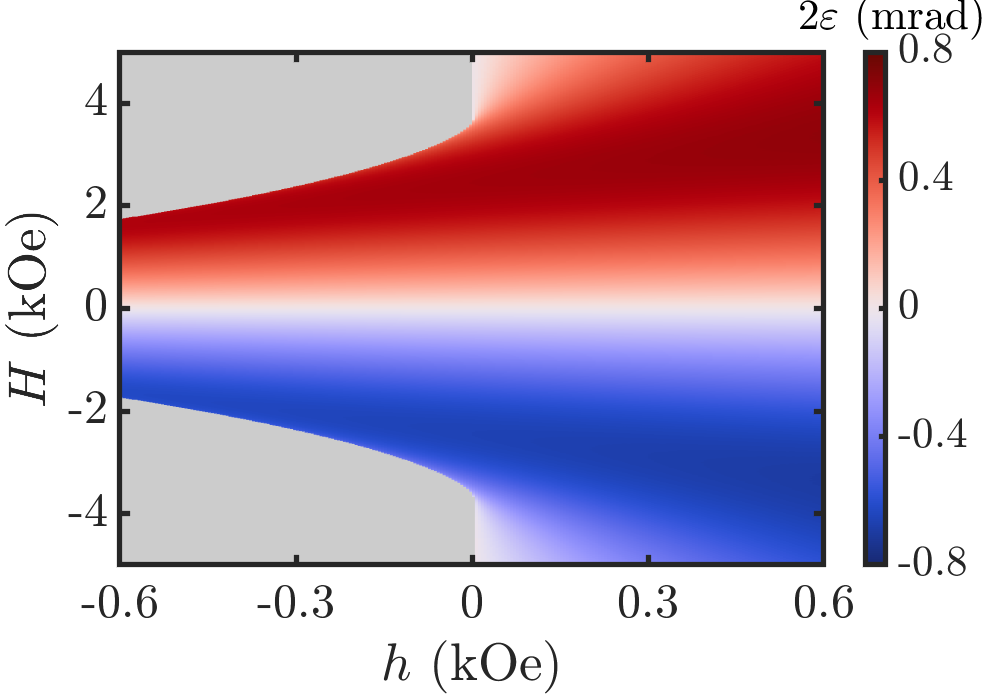}
\includegraphics[width=0.68\columnwidth]{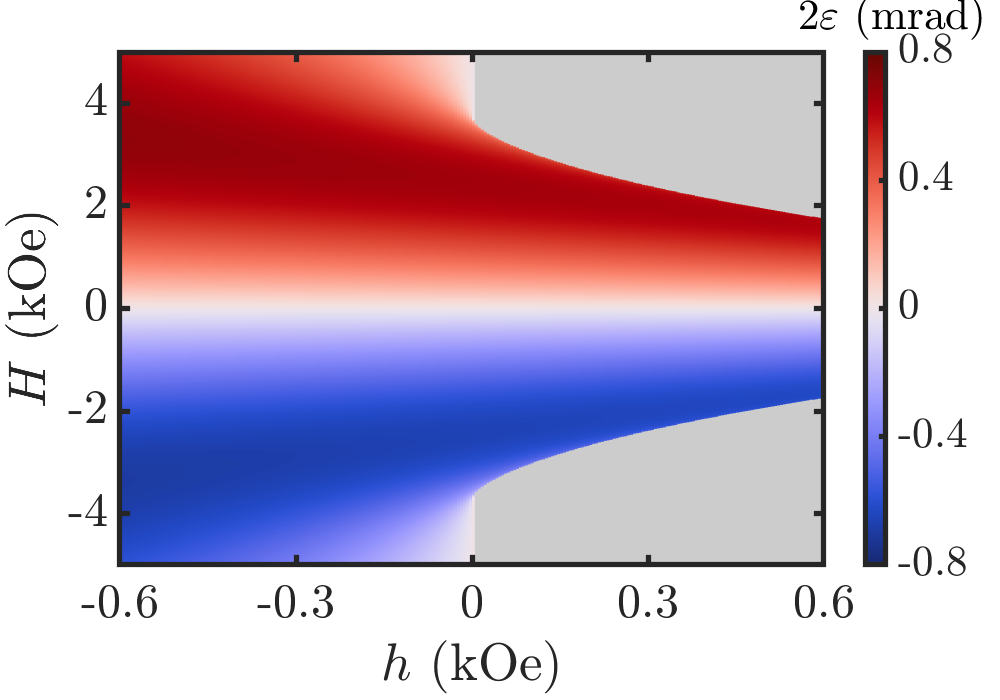}
\includegraphics[width=0.68\columnwidth]{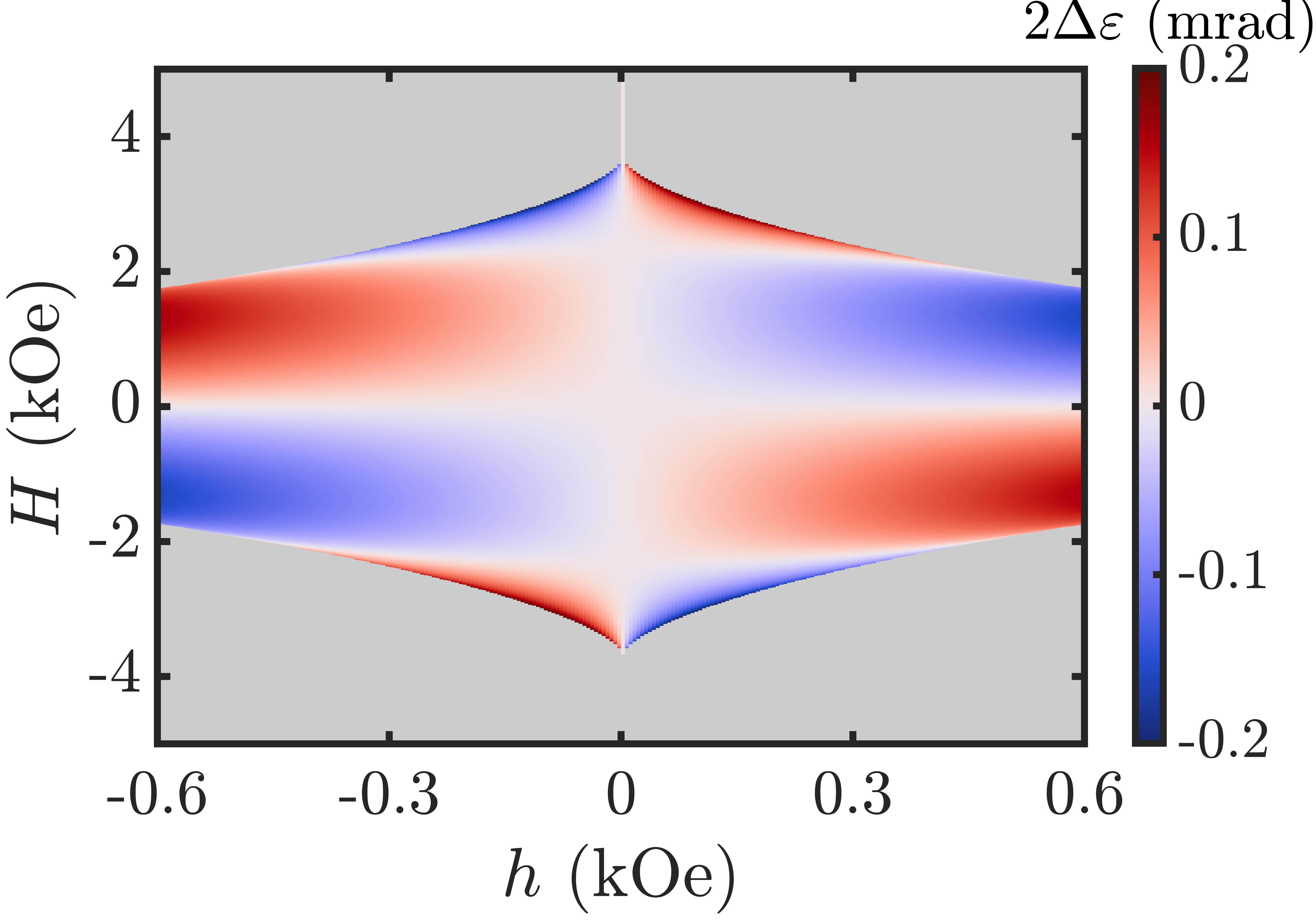}
\caption{Sublattice canting $2\varepsilon$ for (a) $\phi=\pi/2$, and (b) $\phi=3\pi/2$ equilibrium positions in the case of non-zero out-of-plane $h$ field. White area depicts the region where the corresponding solution is abscent. (c) The difference between the sublattice canting $2\Delta \varepsilon = 2\varepsilon_{\pi/2}-2\varepsilon_{3\pi/2}$ vs. in-plane $H$ and out-of-plane $h$ magnetic field components. Grey area shows the region where only one solution exists.}
\label{Fig: eps from hH}
\end{figure*}

Application of the out-of-plane magnetic field results in the disappearance of the true collinear phase. For any finite values of $(H,h)$ the Neel vector $\mathbf{L}$ is tilted with respect to the external magnetic field $\mathbf{H}+\mathbf{h}$, and, as it will be shown further, the two sublattices remain canted. However, the boundary between the bistability region where there are two equilibrium positions of antiferromagnetic vector and the region with a single equilibrium position exists. Moreover, this boundary can be shifted under application of a small out-of-plane magnetic field component $h$. 

There are two ways, how this application of the out-of-plane $h$ field can be implemented in practice. On the one hand, one might fix the position of the sample parallel to the external in-plane magnetic field $\mathbf{H}$ provided by a magnet and apply the out-of-plane field $\mathbf{h}$ using another independent electromagnet. In this case one might tune $H$ and $h$ independently. Fig.~\ref{Fig: phase diagram}a illustrates this case how the boundary of the bistability region shifts for the different values of $h$ field applied. Another way to change the ratio of the in-plane and out-of-plane components is to tilt the sample at an angle $\alpha$ in the external magnetic field $H_0$ created by a single electromagnet. In this case, $H=H_0 \cos \alpha$ and $h=H_0 \sin \alpha$ are simultaneously changed. Fig.~\ref{Fig: phase diagram}b illustrates how the boundary of the bistability region shifts for the different values of a tilt angle $\alpha$ in this case.   

Figures~\ref{Fig: phase diagram}a,b show that application of the out-of-plane $h$ field significantly reduces the bistability region. Quite moderate magnetic fields of tenths of~kOe, or sample tilts of several degrees result in $\sim 1$~kOe shift of the boarder of bistability region. Moreover, the so-called 'waist' of this boarder (the minimal value of $|m|$ in the $m(H)$ dependence describing this border) moves to the region of smaller external magnetic fields of 1-3~kOe which is important for practical realization since can be easily achieved in experimental setups.

Thus, application of small external magnetic fields allows for the efficient control of the boarder of the bistability region, and to perform switching between the stable and quasi-stable states.

\section{Canting of the magnetic sublattices}

It is well-known that in the non-collinear phase under the application of the in-plane magnetic field the sublattices are canted with respect to each other at a small angle $2\varepsilon$, and this canting disappears in the collinear phase. An important question is what is the relative orientation of the magnetization sublattices if out-of-plane magnetic field is applied in addition to the in-plane one.

Considering $\varepsilon, \beta$ as small corrections to $\theta, \phi$ one may obtain the following relations from Eq.~\eqref{Eq Phi gen}:
\begin{eqnarray}
    \varepsilon &= &\frac{\chi}{M_1+M_2} (H \sin \theta_0 - h \cos \theta_0 \sin \phi_0), \\
    \beta &= &\frac{\chi}{M_1+M_2}  h \frac{\cos \phi_0}{\sin \theta_0},
\end{eqnarray}
where $\theta_0$ and $\phi_0$ are determined by Eq.~\eqref{Eq eq_for_teta}, so that $\phi_0=\pi/2$ or $\phi_0=3\pi/2$ and $\beta = 0$ in all range of $h$ and $H$ values. 

On the contrary, sublattice canting $2\varepsilon$ is quite small ($\sim 10^{-3}$~rad), but non-zero in almost all range of $h$ and $H$ values (Fig.~\ref{Fig: eps from hH}). It is interesting that as $\theta$ differs for the two equilibrium states in the bistable regime (see Fig.~\ref{Fig: teta from H}a-c), sublattice canting $2\varepsilon$ is also different for these states (Fig.~\ref{Fig: eps from hH}c). This difference reaches its maximum at the edge of the bistability region.

For the large values of $H$ where according to Eq.~\eqref{Eq delta_teta quasi-collinear} $|\sin \theta|\rightarrow 0$, sublattice canting $2\varepsilon$ gradually decreases $2\varepsilon \rightarrow 0$. However, for any finite value of $H$ it is still non-zero (Fig.~\ref{Fig: eps from hH}a,b). This explains why the second-kind phase transition between the collinear and non-collinear phases disappears in the case of out-of-plane magnetic field applied to the system. According to our analysis, this happens due to the disappearing of the collinear state.

\section{Conclusion}

We report a theoretical study of the phase diagram of a ferrimagnetic iron-garnet with uniaxial anisotropy near a magnetization compensation point in the presence of the two-component magnetic field. The equilibrium state was determined using an effective energy function in the quasi-antiferromagnetic approximation. It was shown that the phase diagram and the Neel vector equilibrium positions change significantly in the presence of the two-component magnetic field, with the components aligned along and perpendicular to the easy axis. Switching between the monostable and bistable states becomes possible by tuning the ratio of two magnetic field components. This opens new possibilities for utilization of ferrimagnets, since the magnetic field could be changed much faster than the temperature.

\begin{acknowledgments}
This work was financially supported by the Russian Science Foundation, project No. 23-62-10024.
\end{acknowledgments}

\bibliography{_Manuscript}

\end{document}